\documentclass[useAMS,usenatbib,a4]{mn2e}

\usepackage{epsfig}
\usepackage{times}
\usepackage{graphicx,bm,amssymb}

\newcommand{\mk}{}
\title[MHD relaxation of AGN ejecta]{Magnetohydrodynamic relaxation of AGN ejecta: radio bubbles in the intracluster medium}
\author[Jonathan Braithwaite]{Jonathan Braithwaite \thanks{E-mail: jonathan@astro.uni-bonn.de} \\Argelander Institut f\"ur Astronomie, Auf dem H\"ugel 71, 53121 Bonn, Germany \& \\Canadian Institute for Theoretical Astrophysics, 60 St. George Street, Toronto ON M5S 3H8, Canada}

\pagerange{\pageref{firstpage}--\pageref{lastpage}}
\begin{document}
\maketitle
\label{firstpage}

\begin{abstract}
X-ray images of galaxy clusters often display underdense bubbles which are apparently inflated by AGN outflow. I consider the evolution of the magnetic field inside such a bubble, using a mixture of analytic and numerical methods. It is found that the field relaxes into an equilibrium filling the entire volume of the bubble. The timescale on which this happens depends critically on the magnetisation and helicity of the outflow as well as on properties of the surrounding ICM. If the outflow is strongly magnetised, the magnetic field undergoes reconnection on a short timescale, magnetic energy being converted into heat whilst the characteristic length scale of the field rises; this process stops when a global equilibrium is reached. The strength of the equilibrium field is determined by the magnetic helicity injected into the bubble by the AGN: if the outflow has a consistent net flux and consequently a large helicity then a global equilibrium will be reached on a short timescale, whereas a low-helicity outflow results in no global equilibrium being reached and at the time of observation reconnection will be ongoing. However, localised flux-tube equilibria will form. If, on the other hand, the outflow is very weakly magnetised, no reconnection occurs and the magnetic field inside the bubble remains small-scale and passive. These results have implications for the internal composition of the bubbles, their interaction with ICM -- in particular to explain how bubbles could move a large distance through the ICM without breaking up -- as well as for the cooling flow problem in general. In addition, reconnection sites in a bubble could be a convenient source of energetic particles, circumventing the problem of synchrotron emitters having a shorter lifetime than the age of the bubble they inhabit.
\end{abstract}

\section{Introduction}\label{sec:intro}

The gravitational potential wells of galaxy clusters are filled with hot ($10^{7-8}$ K), hydrostatically-settled gas which emits X-rays via thermal bremsstrahlung (e.g. \citealt{Molendi:2004}). Many galaxy clusters, viewed in X rays, display dark cavities of size $\sim10$ kpc at various distances from the cluster centre {\mk \citep{Boehringer_etal:1993,Carilli_etal:1994,Dunn_Fabian:2004,McNamara_Nulsen:2007,Birzan_etal:2008}.} They are dark because they have a lower density than the surrounding intra-cluster medium (ICM), but precisely how much less dense is uncertain, except that they are at least a factor of three or so less dense. {\mk Observationally constraining the density is difficult because the line of sight contains also surrounding material; this problem becomes more severe where the bubble is at a larger distance from the cluster centre (see \citealt{Ensslin_Heinz:2002} for details).} The bubbles are apparently inflated by an Active Galactic Nucleus (AGN) at the cluster centre and then rise buoyantly through the ICM. In addition, we infer the presence of an internal magnetic field and cosmic rays from observed radio synchrotron emission. 

There is a growing consensus that negative feedback from AGN could solve the cooling flow problem: the accretion of gas onto a supermassive black hole in the central galaxy releases energy to heat the ICM, preventing it from cooling and collapsing towards the centre of the cluster.  Observationally, there is a strong correlation between those clusters which require heating (i.e. have a short cooling timescale) and the presence of optical-line emission and radio emission from AGN as well as star formation \citep{Burns:1990,Rafferty_etal:2008,Cavagnolo_etal:2008}. The means by which this energy might be transferred to the ICM is not yet understood, but an interaction between the AGN outflow and the surroundings does seem very likely {\mk (e.g. \citealt{Brueggen_Kaiser:2002,Reynolds_etal:2002,Churazov_etal:2005,Brueggen_etal:2005,Brighenti_Mathews:2006}).} For this reason, it is important to gain some understanding of how AGN-inflated bubbles interact with their surroundings.

Rising bubbles in a fluid tend to lose their spherical shape after rising a distance comparable to their radius. First, bubbles tend to flatten while the surrounding medium is flowing around them because the material flowing past their sides is moving with greater velocity than either in front of or behind them; we know from Bernoulli's principle that the pressure at the sides must therefore be lower and so the bubble expands laterally. {\mk Then, the bubble is shredded into many smaller bubbles and eventually becomes completely mixed into the surrounding medium. In general there is more than one instability responsible for this shredding: the Rayleigh-Taylor (R-T) instability appears at the leading edge of the bubble where a dense fluid (the ICM) lies above a less dense fluid (the bubble) and the Kelvin-Helmholtz (K-H) instability appears at the sides of the bubble where there is a discontinuity in velocity and density. In the absence of magnetic fields, the growth time of the longest wavelength mode (i.e. the bubble radius) of the R-T instability is comparable to the time the bubble takes to rise a distance equal to its own size. The growth time of the K-H instability is likely to be somewhat longer if there is a large density contrast between the bubble and its surroundings. However, in many clusters we see large bubbles which have risen distances many times greater than their own size -- some mechanism must be inhibiting the instabilities \citep{JonandDeY:2005,Ruszkowski_etal:2007}.} An obvious candidate is a magnetic field, coherent on the length scale of the bubble, either in the ambient medium (`magnetic draping', see \citealt{Lyutikov:2006,DurandPfr:2008}) or inside the bubble, or both. Alternatively, by analogy with smoke rings it seems plausible that there is some purely hydrodynamical process responsible -- for instance \citet{ScaandBru:2009} and \citet{Brueggen_etal:2009} find that adding a subgrid-turbulence model to hydro simulations could encourage the bubble to stay in one piece. Here, I concentrate on the magnetic field inside the bubble, and show how an arbitrary `turbulent' magnetic field in a new-born bubble could reconnect into a large-scale equilibrium and thus provide the necessary rigidity. This process is similar to that taking place in stars which make a transition from convective to non-convective, for instance in proto-neutron stars \citep{BraandSpr:2004,Braithwaite:2008}.

In section \ref{sec:analytic} I look at the process of relaxation to equilibrium, finding a relation between the initial magnetic helicity and the equilibrium field strength as well as comparing the relevant timescales. In section \ref{sec:sims} I present numerical simulations of the reconnection process, before looking in some detail in section \ref{sec:structure} at the structure of the equilibria found. In sections \ref{sec:disc} and \ref{sec:conc} I discuss the results and then summarise and conclude.

\section{Formation of equilibrium inside a bubble}\label{sec:analytic}

In this section I look at the properties of magnetic bubbles and make estimates of the relevant timescales and energies.

Imagine a bubble of hot gas which, after becoming detached from the AGN contains an initially turbulent, disordered magnetic field. As we shall see below, it is likely that important changes in the magnetic field inside the bubble will happen on a short timescale compared to the bubble's buoyant rise through the ICM, so we shall not consider the interaction with the ICM. Now, generally the field left over from the AGN outflow will not be in equilibrium, meaning that the Lorentz force the magnetic field exerts on the gas is not balanced by the gas pressure gradient. Fluid motion results and kinetic energy is dissipated by viscosity, which in this low-density case is quite high. Eventually the free energy in the magnetic field is used up and a stationary equilibrium is reached.\footnote{Strictly speaking, the equilibrium should continue to evolve due to the finite conductivity of the plasma and the resulting energy loss. We can think of the evolving equilibrium as an electric circuit consisting of an inductance and a resistance, but assuming the standard Spitzer conductivity the timescale for decay of the current and magnetic field is $\sim10^{33}$ yrs.} Before looking at the timescale for relaxation to equilibrium in section~\ref{sec:timescales}, I first calculate the strength of the equilibrium field, making use of the conserved quantity of magnetic helicity.

\subsection{Energy and helicity of the equilbrium}\label{sec:en_hel}

Let us imagine the relaxation to magnetohydrodynamic equilibrium of a bubble with radius $r$, thermal pressure $P$ and density $\rho$ which contains a magnetic field of energy $E=V B^2/8\pi$, where $B$ is the r.m.s.\ magnetic field inside the bubble and $V=4\pi r^3/3$ is the volume of the bubble. After relaxation, an equilibrium is reached. In the following, quantities inside the bubble initially and (finally) at equilibrium are marked with the subscripts i and f respectively, and quantities outside the bubble -- which are assumed not to change during the relaxation to equilibrium -- have the subscript o. We can say the following about the equilibrium state.

The reconnection destroys magnetic energy on small length scales but has little effect on the magnetic helicity, a global quantity which is defined as the volume integral of the scalar product of the magnetic field with its vector potential $H\equiv(1/8\pi)\int\!{\bf A}\!\cdot\!{\bf B}\,{\rm d}V$. It can be shown that in the case of infinite conductivity, helicity is conserved \citep{Woltjer:1958}. Helicity has units of energy times length and so is present more in the larger structures than is the energy -- and it is {\it approximately} conserved during reconnection taking place on small scales, a property which has been very useful in many contexts from the laboratory \citep{ChuandMof:1995,HsuandBel:2002} to the solar corona \citep{ZhaandLow:2003}. Therefore:
\begin{equation}\label{eq:hel_cons}
H_{\rm f} \approx H_{\rm i}.
\end{equation}

\noindent Consideration of dimensions gives us
\begin{equation}\label{eq:hel_val}
|H_{\rm f}| = r_{\rm H,f} E_{\rm f} = \lambda_{\rm f} r_{\rm f} E_{\rm f},
\end{equation}
where $r_{\rm H}$ is the `helicity length' and $\lambda_{\rm f}$ is a dimensionless parameter of the equilibrium whose value can be determined from simulations -- in a large-scale equilibrium it should be of order unity. Having chosen to put $1/8\pi$ into the definition of helicity, we have $r_{\rm H}\sim A/B\sim H/E$.

The thermodynamic relations $e=P/(\gamma-1)$ where $e$ is the internal energy per unit volume and ${\rm d}U=T\,{\rm d}S-P\,{\rm d}V$ give us
\begin{equation}\label{eq:heat}
\frac{P_{\rm f}V_{\rm f}}{\gamma-1} + E_{\rm f} - \frac{P_{\rm i}V_{\rm i}}{\gamma-1} - E_{\rm i} = - P_{\rm o}(V_{\rm f}-V_{\rm i})
\end{equation}
where $\gamma$ is the ratio of specific heats of the bubble gas and $V=4\pi r^3/3$ is the volume of the bubble. It assumed that heat transfer between the bubble and its surroundings can be neglected on the reconnection timescale, that there is no internal source of heat energy, that there is no radiative cooling, so that ${\rm d}S=0$, {\mk and that the reconnection happens sufficiently slowly that there is always pressure balance between the bubble and its surroundings. In fact, (\ref{eq:heat}) is simply an expression of the conservation of enthalpy.} Now, since the bubble is in pressure equilibrium with the surroundings both before and after reconnection, we have
\begin{equation}\label{eq:pressure}
P_{\rm o} = P_{\rm i} + \frac{E_{\rm i}}{3V_{\rm i}} = P_{\rm f} + \frac{E_{\rm f}}{3V_{\rm f}}
\end{equation}
where the coefficients $1/3$ come from the fact that the magnetic field exerts an `isotropic pressure' $P_{\rm mag}\equiv(1/3)B^2/8\pi$, which is equivalent to assigning an adiabatic index of $4/3$ to the magnetic field and using the thermodynamic relation $P=(\gamma-1)e$.

Looking at the equations above, we have four unknowns regarding the final state ($r_{\rm f}$, $P_{\rm f}$, $E_{\rm f}$ and $H_{\rm f}$) and four equations (\ref{eq:hel_cons}) to (\ref{eq:pressure}). [The first equality of (\ref{eq:pressure}) doesn't count, as it relates only quantities in the initial conditions.] So it is now possible to solve for the final state once we know the value of $\lambda_{\rm f}$.

Now, note that if the plasma is relativistic we have $\gamma=4/3$ and $V_{\rm f}=V_{\rm i}$ since both plasma and magnetic field have the same ratio of pressure to energy density, so converting energy from one form to the other has no effect on the total pressure. If on the other hand the fluid is non-relativistic and monatomic, i.e. $\gamma=5/3$, we expect a fractional increase in the bubble's volume; the greatest increase in volume possible occurs if the bubble contains only magnetic energy and no thermal energy at the beginning and where all of this energy is converted to thermal, i.e. where $P_{\rm i}=0$ and $H_{\rm i}=0$, and it is easily found that the fractional increase in volume $V_{\rm f}/V_{\rm i}=8/5$. Making the approximation that $V_{\rm f}\approx V_{\rm i}$, we can retrieve from the above set of equations the following relation
\begin{equation}\label{eq:lambdas}
\frac{B^2_{\rm f}}{B^2_{\rm i}} \approx \frac{\lambda_{\rm i}}{\lambda_{\rm f}},
\end{equation}
where $B$ is the r.m.s.~field strength in the bubble. Note that the final magnetic energy and volume of the bubble are not dependent on the gas density; the latter will certainly affect the time taken to reach equilibrium, and may or may not affect which equilibrium is reached from a given initial magnetic field, but once an equilibrium is reached the density no longer has any effect. I return to the effect of density in section~\ref{sec:sims_density}. If we need to know the density in the equilibrium bubble we have $\rho_{\rm f}/\rho_{\rm i} = V_{\rm i}/V_{\rm f}$, but in any case we see from above that the density can drop by no more than a factor $5/8$.

\subsection{Timescales}\label{sec:timescales}

Imagine an initially stationary fluid of uniform gas pressure containing an arbitrary magnetic field. Looking at the momentum equation and comparing the sizes of the terms, we see that fluid velocities comparable to the Alfv\'en speed will be induced, and so the relaxation to equilibrium must take place at this speed. Generally the equilibrium will be topologically different from the initial conditions and so reconnection of field lines will be required; this reconnection should also take place at the Alfv\'en speed, regardless of the mechanism invoked and its microphysics. Studies of reconnection in various contexts confirm that reconnection does take place at roughly that speed, or rather, somewhat less, say $\alpha v_{\rm A}$ where $\alpha$ has a value of around $0.1$ \citep{Elsner_Lamb:1984,Ikhsanov:2001}. This means that the time required for reconnection to occur across a structure of size $l_{\rm rec}$ is $\tau_{\rm rec}\approx l_{\rm rec}/(\alpha v_{\rm A})$.

Now, imagine that a bubble of radius $r$ contains a magnetic field structured on some length scale $l$. Initially, $l=l_{\rm i}$ and reconnection proceeds on this length scale but as the field relaxes towards equilibrium, the length scale of the magnetic field grows until eventually $l\approx r$. This means that the magnetic field should initially evolve on a timescale of $\tau_{\rm rec}\approx l_{\rm i}/(\alpha v_{\rm A})$ but that as it approaches equilibrium $\tau_{\rm rec}\approx r/(\alpha v_{\rm A})$. At the same time, magnetic energy has been dissipated and $v_{\rm A}$ has fallen, by a large factor if $l_{\rm i}\ll r$. In fact, the magnetic energy should evolve according to
\begin{equation}
\frac{{\rm d}\ln E}{{\rm d}t} \approx -\frac{1}{\tau_{\rm rec}} \approx -\frac{\alpha v_{\rm A}}{l}.
\end{equation}
Clearly the magnetic field evolves slowest when it is closest to a global equilibrium, therefore the time taken to relax to the equilibrium is approximately equal to the reconnection timescale with $l\approx r$:
\begin{eqnarray}\label{eq:recon}
\tau_{\rm relax} \!\!\!\!& \approx & \!\!\!\!\frac{r}{\alpha v_{\rm A}} = \frac{r\sqrt{4\pi \rho}}{\alpha B} = \frac{r}{\alpha}\left(\frac{4\pi r^3 \rho}{6E}\right)^\frac{1}{2}\\
& \approx & \!\!\!\!7.1 \!\times\! 10^6 \left(\frac{\alpha}{0.1}\right)^{-1} \left(\frac{r}{10 {\rm kpc}}\right) \left(\frac{\rho}{{10^{-5}m_{\rm p} \rm cm}^{-3}}\right)^\frac{1}{2} \nonumber\\ & & \left(\frac{B}{20 \mu {\rm G}}\right)^{-1} {\rm yr},
\end{eqnarray}
where $v_{\rm A}$, $B$ and $E$ are the Alfv\'en speed, magnetic field and energy {\it at equilibrium}.

These bubbles are embedded in the hydrostatically-settled intracluster medium and, being less dense, they rise upwards through it. We can calculate the buoyant rise velocity and the associated timescale, which we define as the time taken for the bubble to move a distance equal to its radius, as this is the timescale on which the bubble might be disrupted by instabilities at its surface. Equating the buoyant force to the drag force, we have
\begin{equation}\label{eq:drag}
gV(\rho_{\rm o} - \rho) = \frac{1}{2} \rho_{\rm o} u^2 S C_{\rm d}
\end{equation}
where $g$ is gravity, $\rho_{\rm o}$ is the density of the external medium, $u$ is the terminal velocity, $S$ is the cross-sectional area of the bubble and $C_{\rm d}$ is the drag coefficient, which has a value of around $0.5$ for a solid sphere at Reynolds numbers of $\sim 10^5$. However, \citet{Churazovetal:2001} find a higher effective value in the context under consideration here, owing to extra energy loss from the excitation of internal gravity waves in the ambient medium; below, we shall adopt their value of $C_{\rm d}=0.75$. Also, we can express gravitational acceleration $g$ in terms of the Keplerian velocity $v_{\rm Kep}$ of a circular orbit at distance $R$ from the centre of the cluster, $g=v_{\rm Kep}^2/R$. Assuming the bubble is spherical, the terminal velocity $u$ is given by 
\begin{equation}\label{eq:term_vel}
u^2 \approx \frac{8}{3C_{\rm d}} \, \frac{\rho_{\rm o}-\rho}{\rho_{\rm o}} \, \frac{r}{R} \, v_{\rm Kep}^2.
\end{equation}
Note that the rise velocity is comparable to the Keplerian velocity, which in turn must be comparable to the sound speed in the hydrostatically-supported surrounding medium $c_{\rm s, o}$; the relation between the two is $v_{\rm Kep}^2/R=c_{\rm s, o}^2/(\gamma_{\rm o} H_P)$ where $H_P$ and $\gamma_{\rm o}$ are the pressure scale height and adiabatic index ($=5/3$) of the surrounding gas. However, the motion of the bubble is not likely to be fast enough that significant energy is dissipated in shocks.
The rise timescale is
\begin{eqnarray}\label{eq:risetime}
\tau_{\rm rise}\!\!\!\! & = &\!\!\! \frac{r}{u} \approx \left(\frac{3C_{\rm d}}{8}\right)^{\frac{1}{2}}\left(\frac{R}{r}\right)^{\frac{1}{2}}\frac{r}{v_{\rm Kep}}\\
& \approx &\!\!\!\! 5 \!\times\! 10^6 \left(\frac{R}{r}\right)^\frac{1}{2} \!\left(\frac{r}{10\,{\rm kpc}}\right) \!\left(\frac{v_{\rm Kep}}{1000 \,{\rm km\,s}^{-1}}\right)^{-1} {\rm yr},
\end{eqnarray}
taking $\rho_{\rm o}/(\rho_{\rm o} - \rho)\approx1$, which seems justified by the X-ray observations.
As a check, we can also calculate the time taken for the bubble to reach this terminal velocity $u$ by looking at the acceleration $a$ from rest:
\begin{equation}\label{eq:acc}
\tau_{\rm acc} = \frac{u}{a} = \frac{u\rho^\prime}{g(\rho_{\rm o}-\rho)}
\approx \left(\frac{8}{3C_{\rm d}}\frac{r}{R}\right)^\frac{1}{2} 
\frac{R}{v_{\rm Kep}},
\end{equation}
{\mk where $\rho^\prime$ is some density which accounts for the inertia of the bubble and of the surrounding gas; it is comparable to $\rho_{\rm o}$. Comparing this to the rise timescale at terminal velocity $\tau_{\rm rise}$ we find that $\tau_{\rm acc}/\tau_{\rm rise} \approx 8/3C_{\rm d}$ which is roughly equal to unity.} We therefore make the approximation that the bubble always moves at its terminal velocity.

We can now compare the relaxation timescale $\tau_{\rm relax}$ to the rise timescale. Dividing (\ref{eq:recon}) by (\ref{eq:risetime}) we find that
\begin{eqnarray}\label{eq:timeratio}
\frac{\tau_{\rm relax}}{\tau_{\rm rise}} \!\!&\!\! \approx \!\!&\!\! 
\frac{u}{\alpha v_{\rm A}}\approx\left(\frac{8}{3C_{\rm d}}\right)^\frac{1}{2}\left(\frac{r}{R}\right)^\frac{1}{2}\frac{v_{\rm Kep}}{\alpha v_{\rm A}}\nonumber\\
&\!\! \approx \!\!&\!\! 
\left(\frac{4 R}{9C_{\rm d}H_P}\right)^{\frac{1}{2}}
\left(\frac{r}{R}\right)^\frac{1}{2} \frac{1}{\alpha} \left(\frac{P_{\rm o}\rho}{P_{\rm mag}\rho_{\rm o}}\right)^{\frac{1}{2}} \nonumber\\
\!\!&\!\! \approx & 1.3  \left(\frac{r}{R}\right)^\frac{1}{2}  \left(\frac{\alpha}{0.1}\right)^{-1}\left(\frac{v_{\rm Kep}}{1000 \,{\rm km\,s}^{-1}}\right)  \nonumber\\
\!\!&\!\! \!\!&\left(\frac{\rho}{10^{-5}m_{\rm p}\,{\rm cm}^{-3}}\right)^\frac{1}{2}  \left(\frac{B}{20\,\mu {\rm G}}\right)^{-1}.
\end{eqnarray}
On the second line above, the ratio has been expressed as the product of two factors of order unity in addition to a pressure ratio, a density ratio, and a factor of $1/\alpha$. The quantity $P_{\rm mag}\equiv B^2/24\pi$ is the isotropic magnetic pressure (a quantity which appears again in section~\ref{sec:structure}) inside the bubble. It is these ratios $P_{\rm o}/P_{\rm mag}$ and $\rho/\rho_{\rm o}$ which are most uncertain; it seems likely that the former is rather high and that the latter is rather low. In the third line some likely numerical values are given which produce a ratio of timescales of about $1$. 

In light of these estimates, it is natural to ask what the magnetic field should look like if we observe it before it has reached a global equilibrium. Observationally, there are the following possibilities:
\begin{enumerate}
\item The initial field is weak and no significant reconnection occurs. The dominant length scale in the field $l$ remains at its initial value $l_{\rm i}$ where $\tau_{\rm rec} \approx l_{\rm i}/(\alpha v_{\rm A}) > \tau_{\rm age}$ where $\tau_{\rm age}$ is the age of the bubble. The field evolves passively in response the the interaction between the bubble and its surroundings.
\item The field is stronger and reconnection proceeds initially on a short timescale, but because of low helicity the equilibrium field strength is low and the Alfv\'en speed drops by a large factor as reconnection proceeds. At the time of observation, the field is still structured on scales small compared to the size of the bubble: $l_{\rm i}<l<r$. If it were possible to measure $l$ and $v_{\rm A}$, we would find that $l/(\alpha v_{\rm A})\approx \tau_{\rm age}$.
\item The helicity of the field is large so that the equilibrium energy as calculated in (\ref{eq:hel_val}) is also large. Reconnection proceeds and the length scale of the field grows without the Alfv\'en speed becoming very low. A global equilibrium is reached ($l\approx r$) and we would measure that $r/(\alpha v_{\rm A}) < \tau_{\rm age}$.
\end{enumerate}

Given the large uncertainly in the parameters, it is not at all clear whether the field inside the bubble should have time to reorganise itself into an equilibrium before the bubble rises and is disrupted into a `mushroom-cloud' shape. The magnetic field strength in observed bubbles could well be somewhat greater than $20\mu$G, or the density could be much lower -- indeed the material in the bubble could be a pair plasma instead of ionised hydrogen -- both leading to effective relaxation on a short timescale and possible stabilisation of a bubble against shredding instabilities, but it is impossible at present to say for certain whether the ratio of the two timescales (\ref{eq:timeratio}) in observed systems is less than, equal to, or greater than unity, and it is possible that all three regimes exist in different bubble systems, given the diversity in observational properties between different galaxy clusters.

\subsection{Input from AGN outflow}\label{sec:jet}

We have seen above how the evolution of a bubble depends on the magnetic helicity of the field it contains, as well as the field strength -- it would be useful therefore to look at how much magnetic energy and helicity an AGN is likely to contribute.

The bubbles are presumably inflated by some jet or other outflow from a system consisting of a supermassive black hole and an accretion disc. I assume here that the bubble is inflated by a magnetocentrifugally-accelerated jet from an accretion disc with a net-flux magnetic field (see e.g. \citealt{BlaandPay:1982}, \citealt{Moll:2009}, and refs therein). The net poloidal flux comes from the accreted material and may therefore change in time, but assuming that at any one instant the disc is threaded by magnetic field of a particular direction, the material in the jet will also contain a poloidal field in this direction. The direction of the toroidal field in the jet will also depend on the direction of rotation of the disc, such that the toroidal field will be in the same sense as the rotation of the disc if the poloidal field is directed into the disc and it will be in the opposite sense to the rotation if the poloidal field is directed out of the disc. Since we can think of helicity as the product of poloidal and toroidal fluxes, we see that the two jets, and therefore bubbles, on either side of the disc will have equal and opposite magnetic helicities. This is obviously convenient because the disc does not need to produce any helicity itself; rather, it merely transports helicity between the two hemispheres.

The following is a rudimentary estimation of the magnetic energy and helicity in a bubble. Although AGN jets are thought to have bulk Lorentz factors $\sim 10$, the calculation here does not take relativistic effects into account: while there is no certainty that the bubble is in fact inflated by a relativistic outflow -- it is likely that the majority of mass and magnetic flux comes from a non-relativistic disc wind -- there seems little purpose in conducting a relativistic generalisation at this stage.

The system can be characterised by the half-width of the outflow at the Alfv\'en surface $r_{\rm j}$ as well as the gas pressure $P_{\rm j}$, the density $\rho_{\rm j}$, the flow speed $v_{\rm j}$ and the magnetic field $B_{\rm j}$ at the Alfv\'en surface. At the Alfv\'en surface, the poloidal and toroidal components of the field are roughly equal, and noting that helicity is the product of poloidal and toroidal fluxes we may write down the following expression for the helicity crossing the Alfv\'en surface per unit time:
\begin{equation}
H_t \sim \Phi_{\rm pol} \Phi_{{\rm tor}, t} \sim B_{\rm j}^2 r_{\rm j}^3 v_{\rm j}
\end{equation}
where the subscript $t$ denotes a time derivative. All factors of order unity are dropped. The helicity per unit mass injected into the bubble is this quantity divided by the mass injection per unit time $\rho_{\rm j} r_{\rm j}^2v_{\rm j}$, so that the total helicity injected into a bubble of radius $r_{\rm b}$ and density $\rho_{\rm b}$ (using the findings of section \ref{sec:en_hel} that $r_{\rm i}\approx r_{\rm f}$ and $\rho_{\rm i}\approx \rho_{\rm f}$ and using the subscript b for `bubble') is
\begin{equation}\label{eq:est_helicity}
H \sim B_{\rm j}^2 r_{\rm j} r_{\rm b}^3 \frac{\rho_{\rm b}}{\rho_{\rm j}}.
\end{equation}
If there is no net flux through the disc, or if the direction of the net flux changes polarity from time to time, the bubble will contain very much less helicity than this estimate, leading to an even weaker field inside the bubble once reconnection has taken place. Incidentally, the jet or outflow will presumably go through a shock as it passes into the bubble -- I make the approximation here that the shock does not dissipate significant magnetic energy or helicity. Imagining that the bubble inflates without any conversion of magnetic energy to heat or vice versa, so that the magnetic field expands adiabatically:\footnote{Strictly speaking, I have also assumed that the expansion is isotropic. Whilst this is clearly not the case in a jet, I am assuming that the expansion on the other side of the shock cancels this effect. The justification for this is the tendency of any magnetised volume to adjust its shape until the three components of the field are roughly equal since this represents the minimum energy (see section \ref{sec:tubes}). Since the three components are roughly equal at the Alfv\'en surface, the total expansion must be the same in each dimension.}
\begin{equation}\label{eq:est_Bi}
\left(\frac{B_{\rm i}}{B_{\rm j}}\right)^2\sim\left(\frac{\rho_{\rm b}}{\rho_{\rm j}}\right)^{4/3},
\end{equation}
where $B_{\rm i}$ is the field strength in the bubble after inflation but before reconnection has begun. Of course, in reality reconnection will begin whilst the bubble is still inflating, so that the prediction here of whether the field strength is high enough for reconnection to proceed at all can be considered rather conservative.

Now, once inflated the bubble field relaxes into an equilibrium with a dimensionless helicity length $\lambda_{\rm f}\approx1$ whose field strength $B_{\rm f}$ is given by
\begin{equation}\label{eq:est_Bf}
B_{\rm f}^2 \sim  H r_{\rm b}^{-4} \;\;\;\;\;\Rightarrow\;\;\;\;\;\;\;
\left(\frac{B_{\rm f}}{B_{\rm j}}\right)^2 \sim \frac{r_{\rm j}}{r_{\rm b}} \frac{\rho_{\rm b}}{\rho_{\rm j}},
\end{equation}
where (\ref{eq:est_helicity}) has been used. 

The bubble moves away from its `mooring' when the expansion speed ${\rm d}r_{\rm b}/{\rm d}t$ slows down to approximately the buoyant rise speed $u$, which using (\ref{eq:term_vel}) and considering the mass flux into the bubble $\rho_{\rm j}v_{\rm j}r_{\rm j}^2$, gives
\begin{equation}\label{eq:rho_ratio}
\frac{\rho_{\rm b}}{\rho_{\rm j}}\sim\frac{v_{\rm j}}{v_{\rm Kep}}\left(\frac{r_{\rm j}}{r_{\rm b}}\right)^2,
\end{equation}
which can be used to eliminate the ratio $\rho_{\rm b}/\rho_{\rm j}$ in (\ref{eq:est_Bi}) and (\ref{eq:est_Bf}), replacing it with the ratio $v_{\rm j}/v_{\rm Kep}$ whose value is better constrained. Dividing (\ref{eq:est_Bf}) by (\ref{eq:est_Bi}) and using (\ref{eq:lambdas}) and (\ref{eq:rho_ratio}), taking $\lambda_{\rm f}\approx1$, gives
\begin{equation}\label{eq:lambda_i}
\lambda_{\rm i} \sim \frac{r_{\rm j}}{r_{\rm b}}\left(\frac{\rho_{\rm j}}{\rho_{\rm b}}\right)^{1/3}
\sim \left(\frac{r_{\rm j}}{r_{\rm b}}\frac{v_{\rm Kep}}{v_{\rm j}}\right)^{1/3}.
\end{equation}
In a typical AGN/bubble system, we might expect $r_{\rm j}\approx 10^{16}$ cm, $r_{\rm b}\approx 10^{22}\,$cm, $v_{\rm j}\approx c$ and $v_{\rm Kep}\approx c/300$, which gives $\lambda_{\rm i}\sim10^{-3}$. Therefore we expect the energy of the magnetic field to fall by a large fraction during relaxation to equilibrium. The length scale of the field in the bubble before reconnection begins is given by
\begin{equation}\label{eq:l_0}
\frac{l_{\rm i}}{r_{\rm j}} \sim \left(\frac{\rho_{\rm j}}{\rho_{\rm b}}\right)^{1/3} \sim \left(\frac{v_{\rm Kep}}{v_{\rm j}}\right)^{1/3}\left(\frac{r_{\rm b}}{r_{\rm j}}\right)^{2/3}
\end{equation}
from consideration of the expansion of material from the outflow and using (\ref{eq:rho_ratio}). This is consistent with $H\sim E_{\rm i}l_{\rm i}$ which is a special case of the general inequality $H\lesssim El$ (or alternatively $\lambda\lesssim l/r$) in the case where the twist is consistently in one direction, like a box full of right-handed screws.

Defining a magnetisation parameter
\begin{equation}
m\equiv\frac{B_{\rm j}^2}{\rho_{\rm j}c^2},
\end{equation}
it is possible to express the timescale ratio (\ref{eq:timeratio}) in the following way (again, ignoring factors of order unity):
\begin{equation}\label{eq:time_1}
\frac{\tau_{\rm relax}}{\tau_{\rm rise}} \sim 
\frac{1}{\alpha\sqrt{m}}\left(\frac{v_{\rm Kep}}{c}\right)^{7/6}\left(\frac{r_{\rm b}}{r_{\rm j}}\right)^{1/3}.
\end{equation} 
If the outflow is modestly relativistic with $m\approx 10$, we have a timescale ratio of around $10$ with these parameters. Given the approximate nature of this calculation and the uncertainty in the parameters, we cannot therefore say with confidence whether the magnetic field should reorganise into global equilibrium before the bubble moves far. However, we can estimate the initial reconnection timescale $\tau_{\rm rec,i}$:
\begin{equation}
\frac{\tau_{\rm rec,i}}{\tau_{\rm relax}}\approx \frac{(l_{\rm i}/\alpha v_{\rm A,f})}{(r_{\rm b}/\alpha v_{\rm A,i})}\approx \frac{l_{\rm i}}{r_{\rm b}}\frac{B_{\rm f}}{B_{\rm i}}\approx \left(\frac{r_{\rm j}}{r_{\rm b}}\frac{v_{\rm Kep}}{v_{\rm j}}\right)^{1/2},
\end{equation}
using (\ref{eq:l_0}), (\ref{eq:lambdas}) and (\ref{eq:lambda_i}). Since this has a value $<10^{-4}$ and therefore $\tau_{\rm rec,i}/\tau_{\rm rise}<10^{-3}$, we should certainly expect to see reconnection in progress if an equilibrium has not already formed. In terms of the three cases described in section \ref{sec:timescales}, case (i) looks very unlikely unless the outflow is extremely weakly magnetised, in which case it must be driven by some non-magnetic mechanism.

\section{Numerical simulations}
\label{sec:sims}

In this section, simulations of the relaxation of a turbulent field into equilibrium are described.

\subsection{Numerical scheme}

The code used is the {\sc stagger code} (\citealt{NorandGal:1995}, \citealt{GudandNor:2005}), a high-order finite-difference Cartesian MHD code which uses a `hyper-diffusion' scheme, a system whereby diffusivities are scaled with the length scales present so that badly resolved structure near the Nyquist spatial frequency is damped whilst preserving well-resolved structure on longer length scales. This, and the high-order spatial interpolation and derivatives (sixth order) and time-stepping (third order) increase efficiency by giving a low effective diffusivity at modest resolution ($144^3$ here). The code includes Ohmic and well as thermal and kinetic diffusion, which are kept at a low level. The code uses Cartesian coordinates with periodic boundaries, although the computational box is made sufficiently large that nothing significant is happening at the boundaries.

\subsection{Numerical setup and initial conditions}

The bubble is modelled as a sphere of initial radius $r_{\rm i}$ containing hot gas (with an ideal gas equation of state and $\gamma=5/3$) and a turbulent magnetic field, surrounded by a cooler, unmagnetised gas with the same equation of state.

The size of the computational box is chosen so that there are no issues with the bubble material expanding across a boundary and back into itself; it is found that using a computational box of size $6r_{\rm i}$ suffices. In these calculations, there is no gravity, and the ambient medium has a uniform density as we are interested primarily in processes happening on timescales shorter than any buoyancy timescale.

At this stage, we assume the ambient medium is unmagnetised. While this is strictly speaking probably not the case with the radio bubbles observed, it is safe to assume that the Alfv\'en speed inside the bubble is significantly greater than that outside and that therefore during the timescales of interest here, the ambient medium will not evolve. Also, it is possible that the ICM was originally not significantly magnetised and that it became magnetised (and chemically enriched) by galactic mass ejection, whether from AGN or from stars.

The bubble is given an initially random magnetic field which contains energy at a range of length scales: the minimum wavenumber in the initial field is $k_{\rm min}=2\pi/r_{\rm i}$, i.e. the largest length scale present is equal to the bubble radius, and the energy declines to higher wavenumbers as $E(k)\delta k\propto k^{-5/3}$ (for want of anything better).

From the considerations in section~\ref{sec:en_hel}, we have the following degrees of freedom in the initial conditions once we have arbitrarily fixed $r_{\rm i}$ and $P_{\rm o}$ and chosen $\gamma$: an initial helicity parameter $\lambda_{\rm i}=H_{\rm i}/(r_{\rm i}E_{\rm i})$, the density ratio $\rho_{\rm i}/\rho_{\rm o}$ and the pressure ratio $P_{\rm mag}/P_{\rm o}$. In the simulations described below, a range of these parameters is explored. In section~\ref{sec:results_hel}, simulations are run with initial ratios $\rho_{\rm i}/\rho_{\rm o}=1/10$ (consistent with observations) and $P_{\rm mag}/P_{\rm o}=1/2$ and the effect of the magnetic helicity is examined. Later, other parameters are looked at.

\subsection{Dependence on initial helicity}\label{sec:results_hel}

\begin{figure}
\includegraphics[width=1.0\hsize,angle=0]{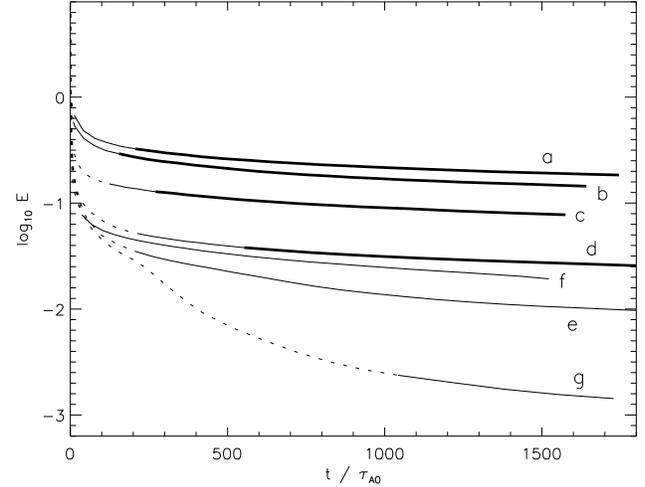}
\caption{Log magnetic energy against time, the former in units of $P_{\rm o}r^3_{\rm i}$ and the latter in units of the Alfv\'en timescale at $t=0$, i.e. $\tau_{{\rm A}0}=r_{\rm i}\sqrt{\rho_{\rm i}V_{\rm i}/(2E_{\rm i})}$, for a set of simulations with various different values of the helicity parameter $\lambda_{\rm i}$. Where the lines are dotted the field is non-equilibrium, solid lines signify some kind of non-simple equilibrium and thick solid lines signify that the field is in a simple axisymmetric equilibrium. The initial energy is the same in all cases ($E_{\rm i}=2\pi$ in these units) but clearly the eventual energy varies between the simulations.}\label{fig:fid-en-t}
\end{figure}

\begin{figure}
\includegraphics[width=1.0\hsize,angle=0]{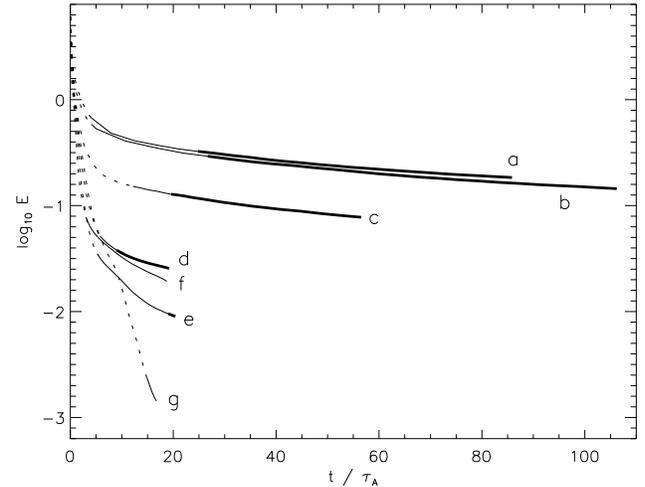}
\caption{As fig.~\ref{fig:fid-en-t} but the time axis has been normalised to give the true number of Alfv\'en timescales which has elapsed. The Alfv\'en timescale increases as the magnetic energy falls so that the time unit in the plot is increasing to the right. In this plot we can see that equilibrium is indeed reached on the order of $\sim10\,\tau_{\rm A}$.}\label{fig:fid-en-t_a}
\end{figure}

\begin{figure}
\includegraphics[width=1.0\hsize,angle=0]{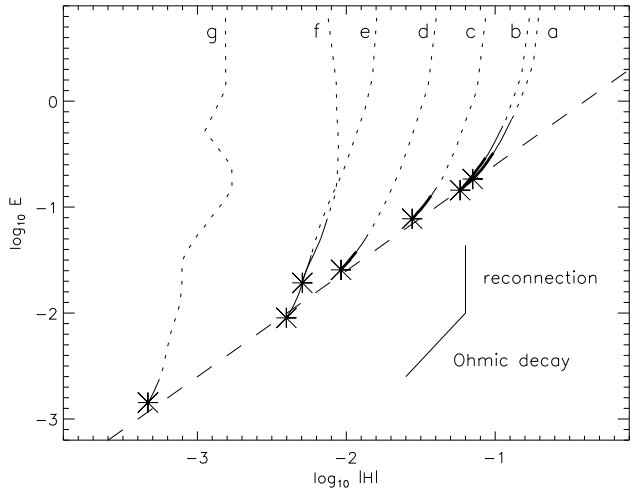}
\caption{Log magnetic energy against log helicity for the same set of simulations as in fig.~\ref{fig:fid-en-t}. Energy and helicity are plotted in units of $P_{\rm o}r^3_{\rm i}$ and $P_{\rm o}r^4_{\rm i}$. As in fig.~\ref{fig:fid-en-t}, where the lines are dotted the field is non-equilibrium, solid lines signify some kind of non-simple equilibrium and thick solid lines signify that the field is in a simple axisymmetric equilibrium. The stars show the position at the end of the run. Helicity falls only a little during the approach to equilibrium, and there is a clear correlation between helicity and the energy at which the equilibrium is reached -- the straight dashed line illustrates the relation $H=0.4r_{\rm i}E$. The solid lines in the lower-right show the gradients at which the field should move during reconnection (which conserves helicity) and during pure Ohmic dissipation ($H\propto E^{2/3}$).}\label{fig:fid-en-hel}
\end{figure}

In this section, a set of simulations with different values of the dimensionless helicity parameter $\lambda_{\rm i}$ is presented. As the field evolves on the dynamical Alfv\'en timescale, we see how the magnetic energy $E$, helicity $H$ and other parameters change. The evolution of energy and helicity in these simulations is plotted in figs.~\ref{fig:fid-en-t}, \ref{fig:fid-en-t_a} and \ref{fig:fid-en-hel}. {\mk The simulations are labelled with letters a-g in the figures.} All simulations were run for the same number of timesteps, which corresponds approximately to the same number (about 600) of sound-crossing times $r_{\rm i}/c_{\rm o}$ where $c_{\rm o}$ is the sound speed in the external medium. Clearly, whilst the energy falls dramatically at the beginning of each run, the helicity falls much more modestly. In fact at first helicity does not seem to fall at all while the energy drops by a factor of ten or more. Then, it is found that in some cases, a simple equilibrium is reached after a short time whereas in other cases the field loses more energy and any equilibrium is often more complicated in shape -- helicity is the determining factor, as expected. In cases with high helicity, less energy is lost and an equilibrium is reached more quickly. It seems though that in all cases some equilibrium is eventually reached, after a number of Alfv\'en crossing times (see fig.~\ref{fig:fid-en-t_a}), the difference being that in some runs the Alfv\'en crossing timescale becomes very long before equilibrium is reached. Also, we see that at equilibrium, all runs lie close to the straight dashed line in fig.~\ref{fig:fid-en-hel}: $|H|=0.4r_{\rm i}E$, which reflects the fact that all equilibria have comparable helicity lengths $|\lambda_{\rm f}|r_{\rm f}\approx0.4r_{\rm i}$, defined in (\ref{eq:hel_val}).

In addition, note that in fig.~\ref{fig:fid-en-hel} it can be seen that once an equilibrium is reached, energy and helicity fall together. This is an effect of finite conductivity; energy should still fall somewhat faster than helicity since the length scale of the equilibrium rises; the equilibrium `spreads out'. To be more precise, we see from integrating over volume the zero-velocity diffusion equation $\partial {\bf B}/\partial t = \eta{\nabla}^2 {\bf B}$ and using Gauss' theorem to equate the right-hand side to zero that $BV\sim Br^3\sim$ const, so that $E\propto r^{-3}$. Now, since $H\sim rE$ we have $H\propto E^{2/3}$ as the equilibrium diffuses (illustrated in fig.~\ref{fig:fid-en-hel}). Also note that flux $\Phi\propto H^{1/2}$.

As the magnetic field evolves, it was found in section \ref{sec:en_hel} that the volume of the bubble increases, and in this case it should increase by a factor $13/10$ if the magnetic energy is largely converted into thermal. In the simulations, some increase in bubble volume is seen, although it is not possible to draw firm conclusions as to whether the volume increase is just that predicted or whether there is an additional increase from mixing of the ambient medium into the bubble. This topic will be explored in a forthcoming publication.

\begin{figure*}
\includegraphics[width=0.288\hsize,angle=0]{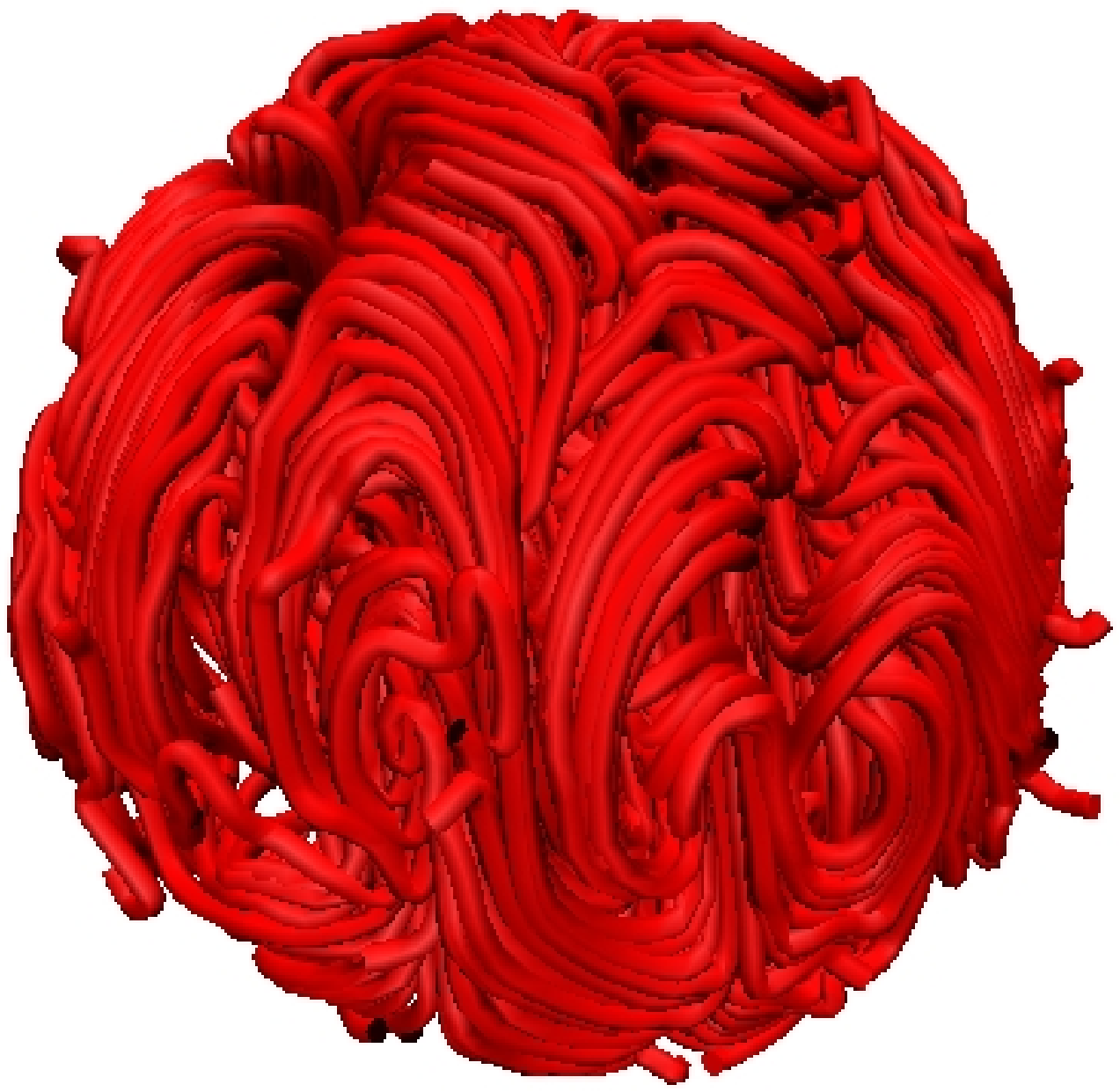}
\includegraphics[width=0.356\hsize,angle=0]{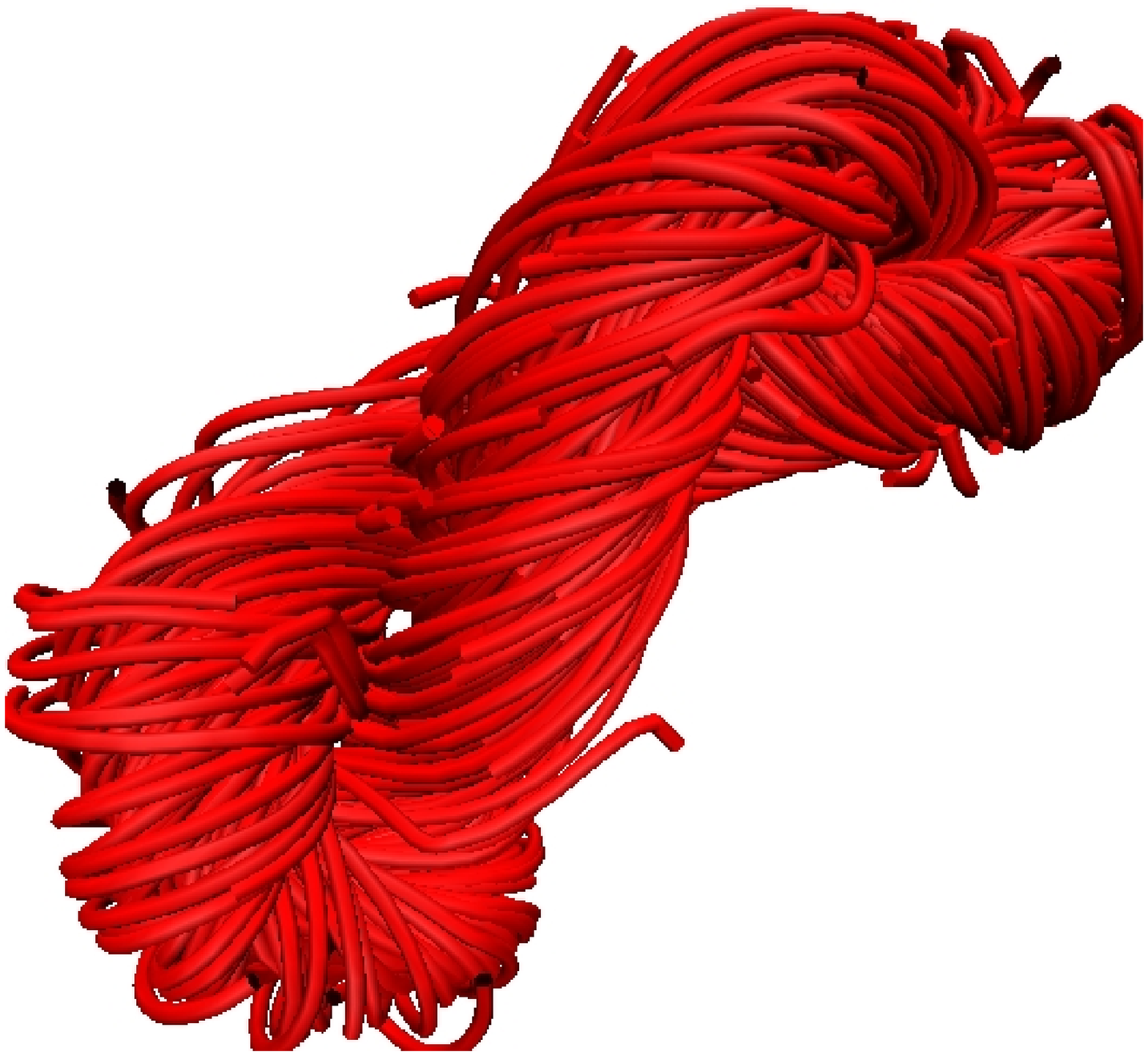}
\includegraphics[width=0.347\hsize,angle=0]{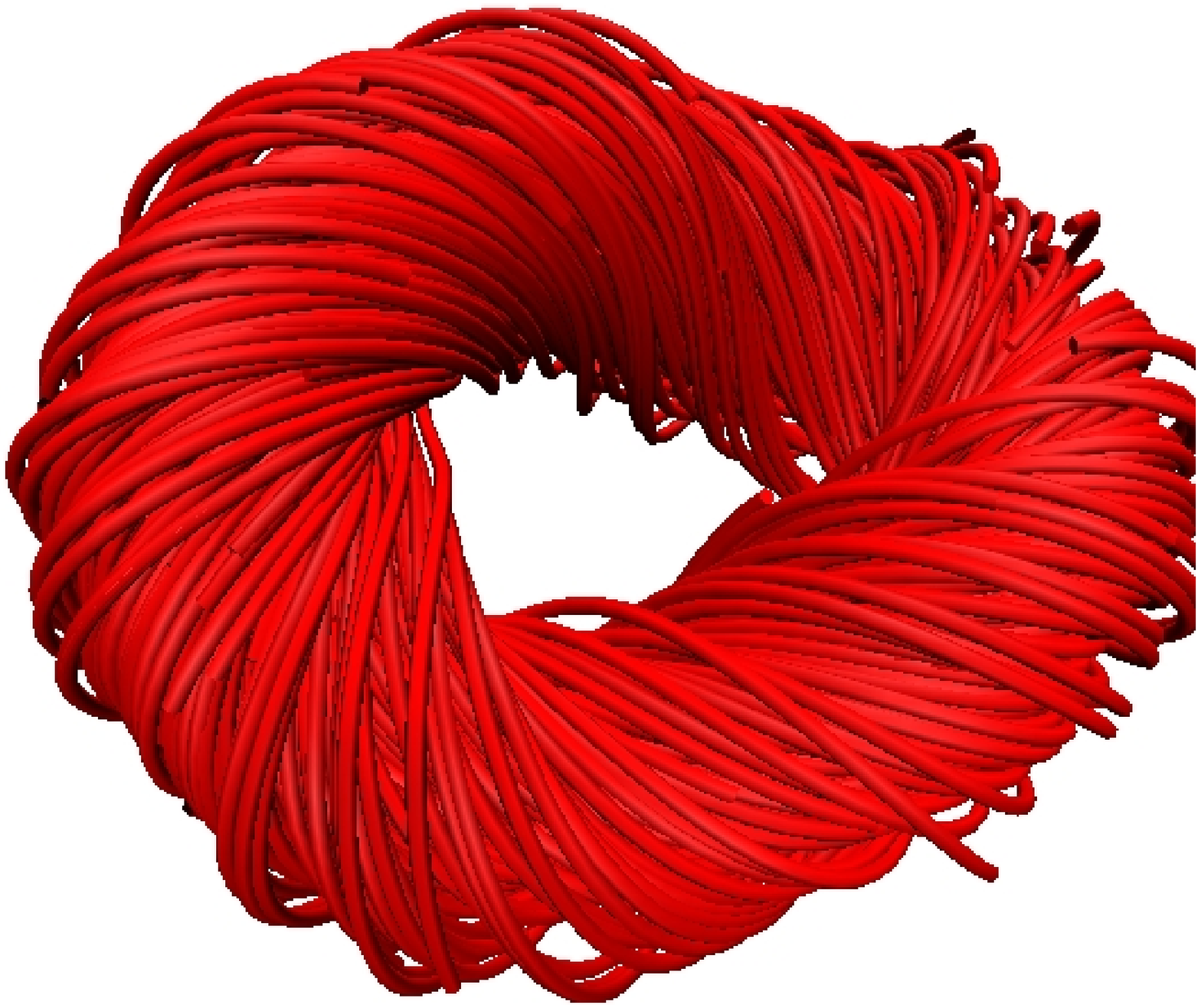}
\caption{Time sequence from the {\mk highest helicity run (a) of those in figs.~\ref{fig:fid-en-t} to \ref{fig:fid-en-hel}.} {\it Left}: initial conditions. {\it Centre}: figure-of-eight configuration. The field can be thought of as consisting of a single flux tube twisted around itself in the opposite direction to its internal twist. {\it Right}: later, once a (roughly) axisymmetric equilibrium is reached: the tube has become a circle. Such an equilibrium can have either negative or positive helicity, corresponding to a clockwise or anticlockwise twist in the field lines.}
\label{fig:vapor1}\end{figure*}

In fig.\ \ref{fig:vapor1} the evolution of the magnetic field in {\mk one of these simulations (the one marked `a') is illustrated. The field has a relatively high helicity ($\lambda_{\rm i}=0.032$) and} the field quickly evolves into a simple torus equilibrium, via a figure-of-eight shaped configuration consisting of a twisted flux tube twisted around itself.

\begin{figure*}
\parbox[t]{0.3\hsize}{\caption{Time sequence {\mk from the second-lowest-helicity run (f)}
, starting from top-middle, then top-right, then bottom left, etc. Two tori form, connected by two thin flux tubes. {\it Bottom-left:} the flat surface is coloured according to field strength: red is weak, green/blue is strong. {\it Bottom-centre:} the tension in the flux tubes pulls the tori together, and a small region of high current density is visible, represented by green volume rendering. {\it Bottom-right:} the two tori have merged. They had opposite sign helicity, reflected in the opposite sense of the twist; the resultant torus, which has much lower energy than either of the two original tori, has the same sense twist as the larger of the two.
\label{fig:merger}}}\,\,\,\,\,\parbox[t]{0.68\hsize}{\mbox{}\\
\includegraphics[width=0.43\hsize,angle=0]{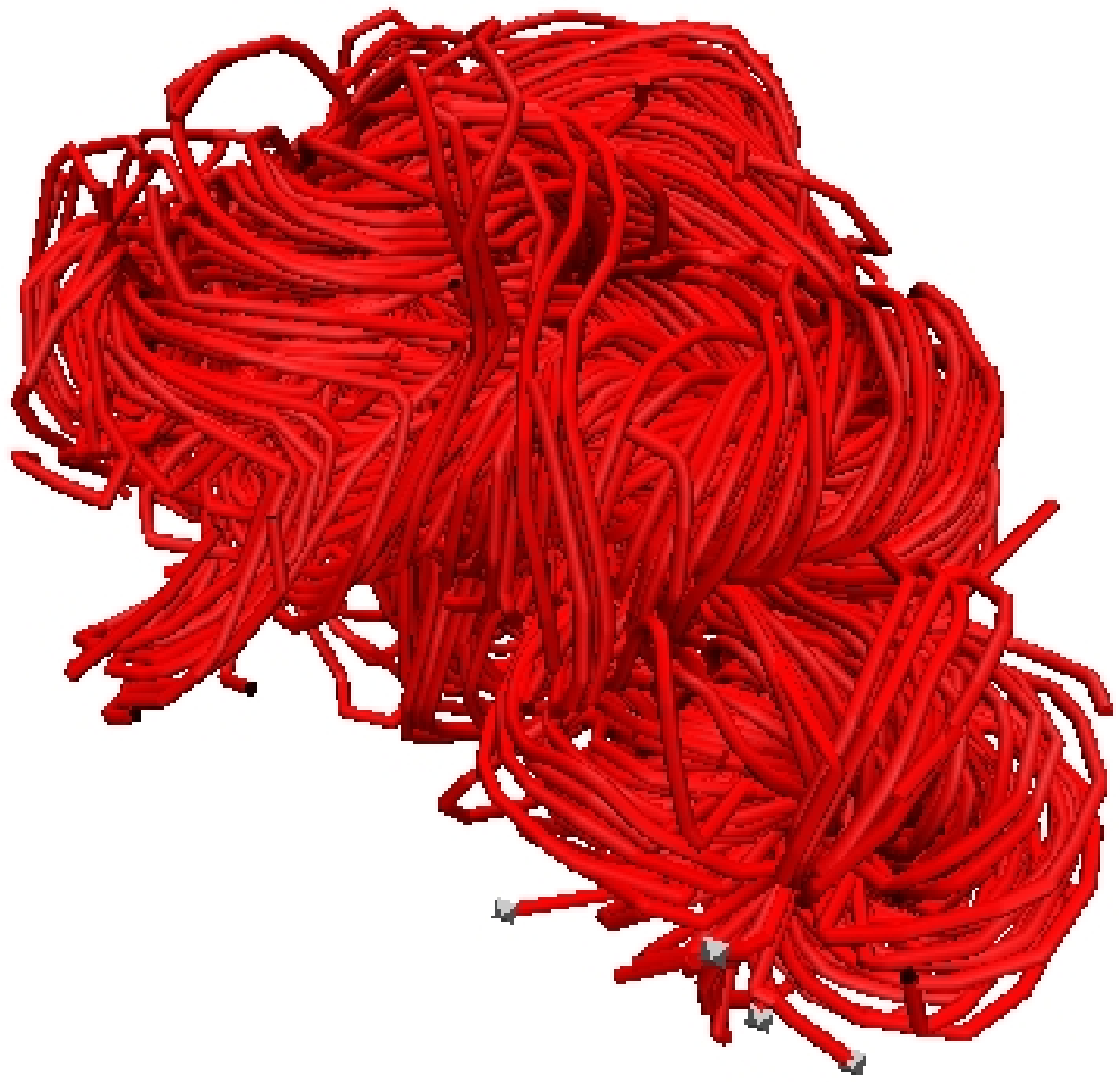}
\includegraphics[width=0.55\hsize,angle=0]{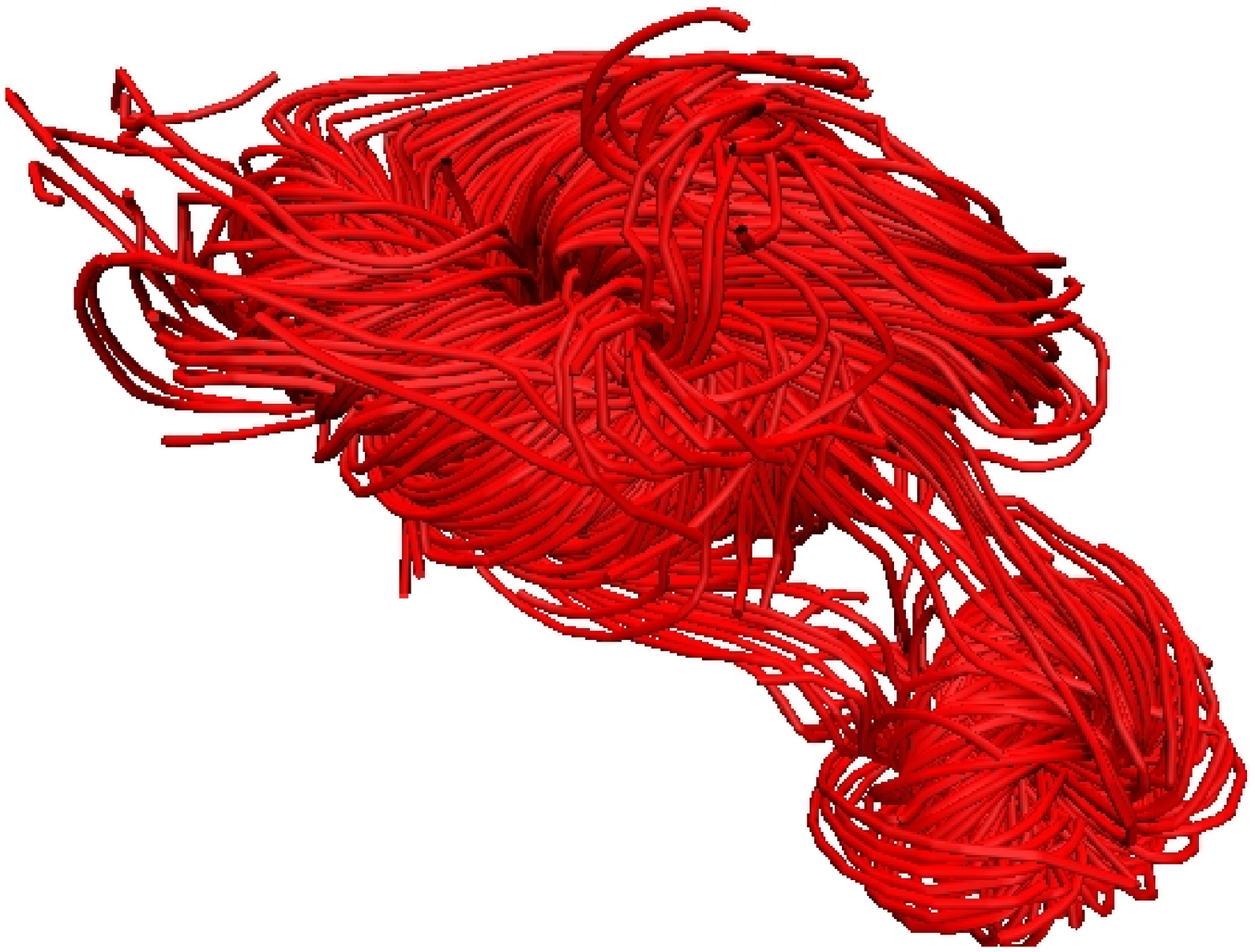}}
\includegraphics[width=0.33\hsize,angle=0]{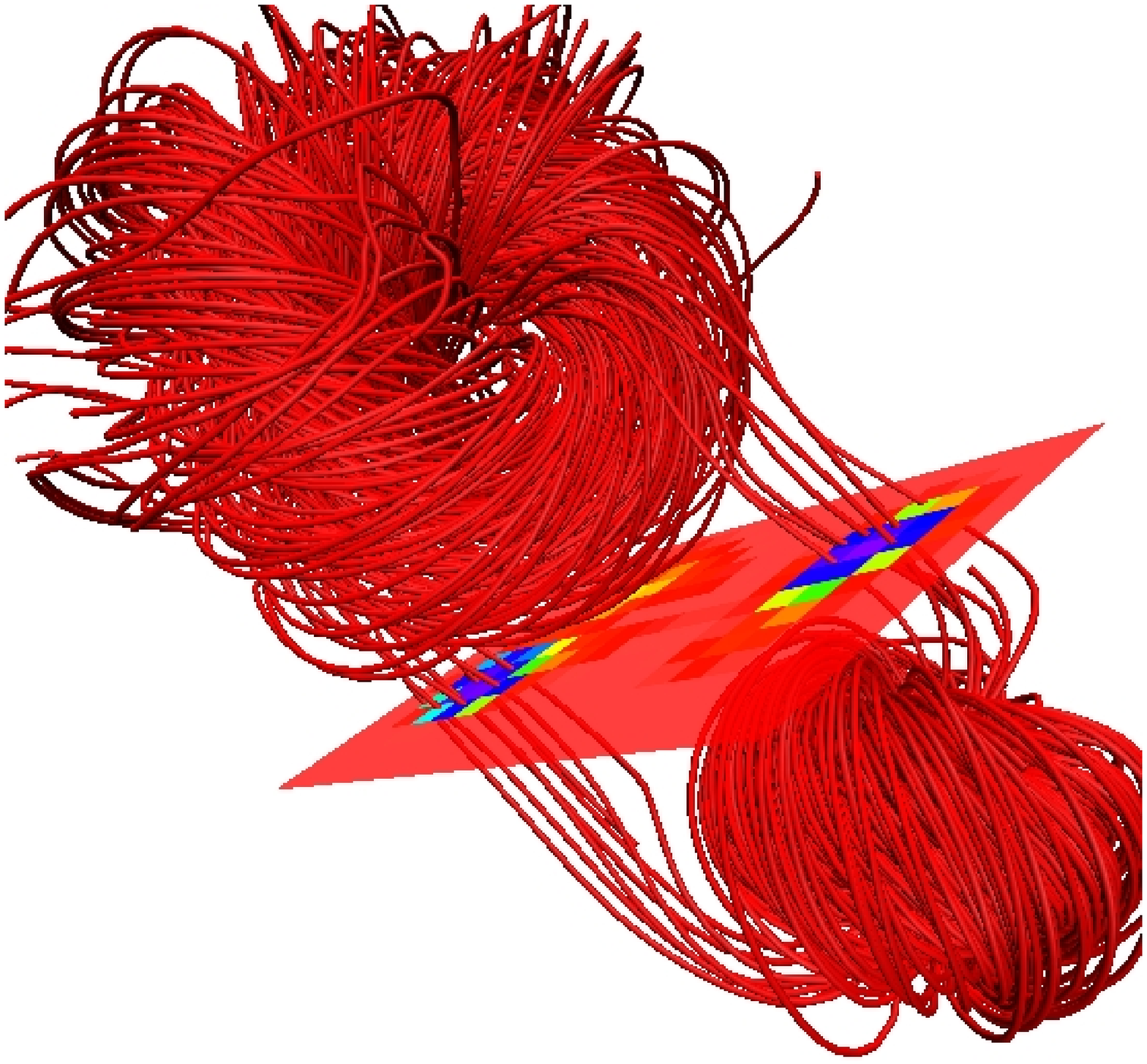}
\includegraphics[width=0.33\hsize,angle=0]{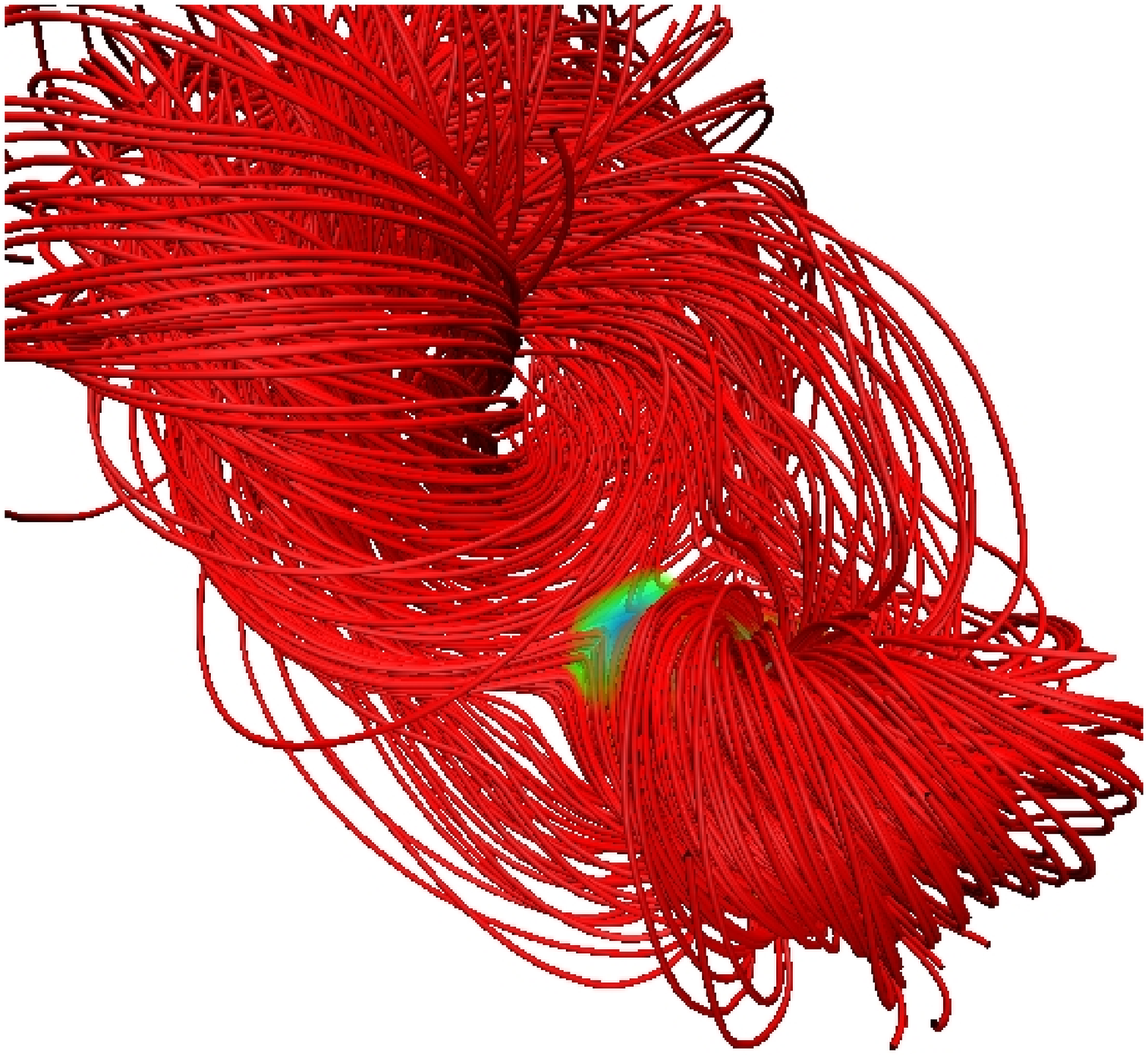}
\includegraphics[width=0.33\hsize,angle=0]{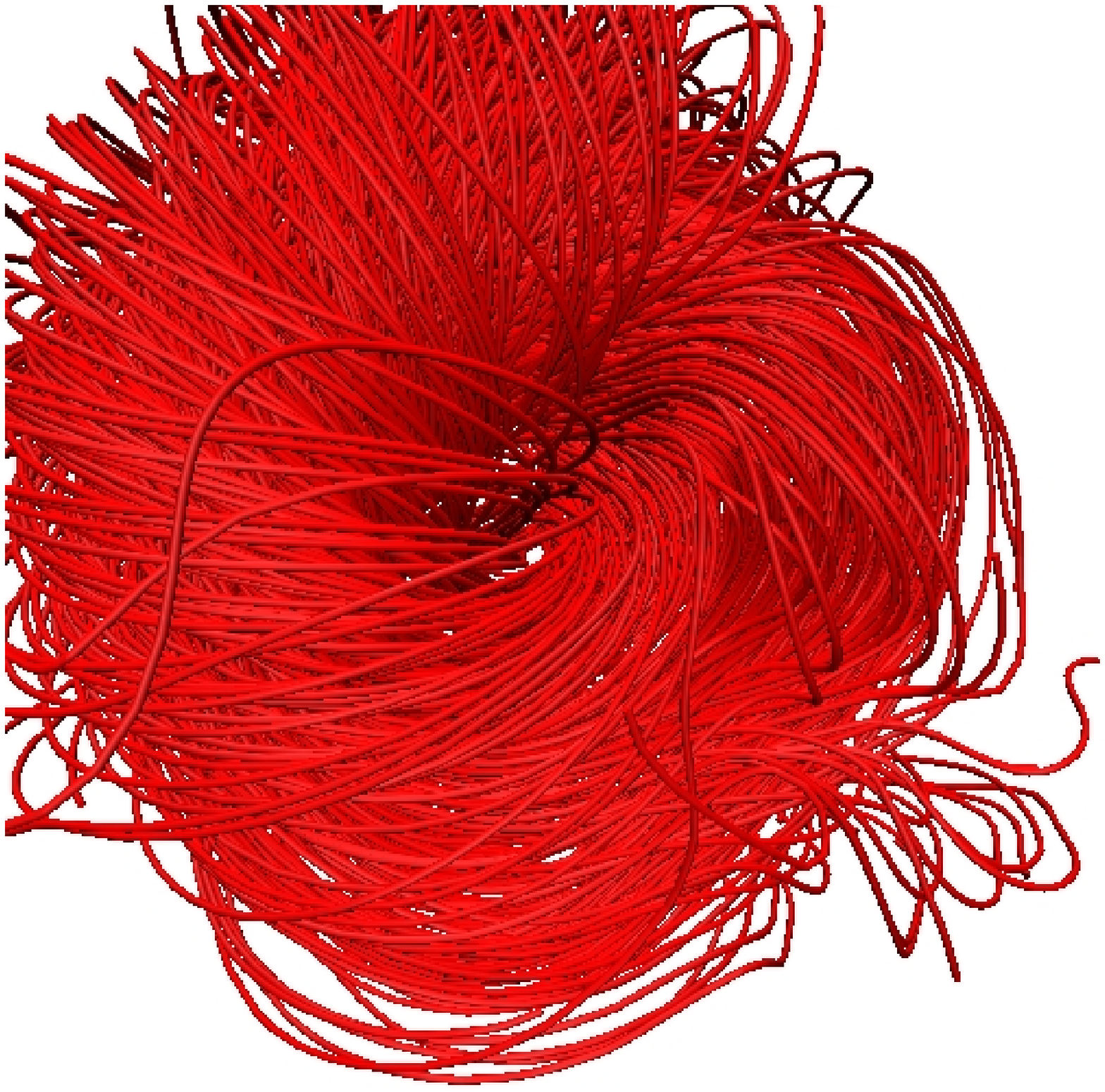}
\end{figure*}

In fig.~\ref{fig:merger} the evolution of the magnetic field in {\mk another simulation (f) is illustrated, this time with lower helicity parameter $\lambda_{\rm i}=0.0013$.} An intermediate state is reached which consists of two torus-shaped fields connected by two flux tubes; however the two tori are pulled together again by the tension in the tubes which join them, and a current sheet forms at the interface. Eventually the two tori become one, significantly weaker torus.

It appears that the most basic equilibrium is a `ball of string' twisted torus shape. All equilibria consist of twisted flux tubes arranged in some pattern, the simple torus being a special case where the flux tube makes a circle. Tubes may be twisted in either sense, corresponding to positive and negative magnetic helicity. Magnetic fields with greater helicity tend to evolve directly into simpler equilibria; conversely when the helicity is very small the energy drops by a large factor, reducing the Alfv\'en speed to such an extent that continued simulation of the evolution becomes impossible. It seems plausible that all fields eventually evolve into a simple torus configuration, the important question being whether this happens on a sufficiently short timescale. 

\subsection{Density of the bubble}\label{sec:sims_density}

\begin{figure*}
\includegraphics[width=0.33 \hsize,angle=0]{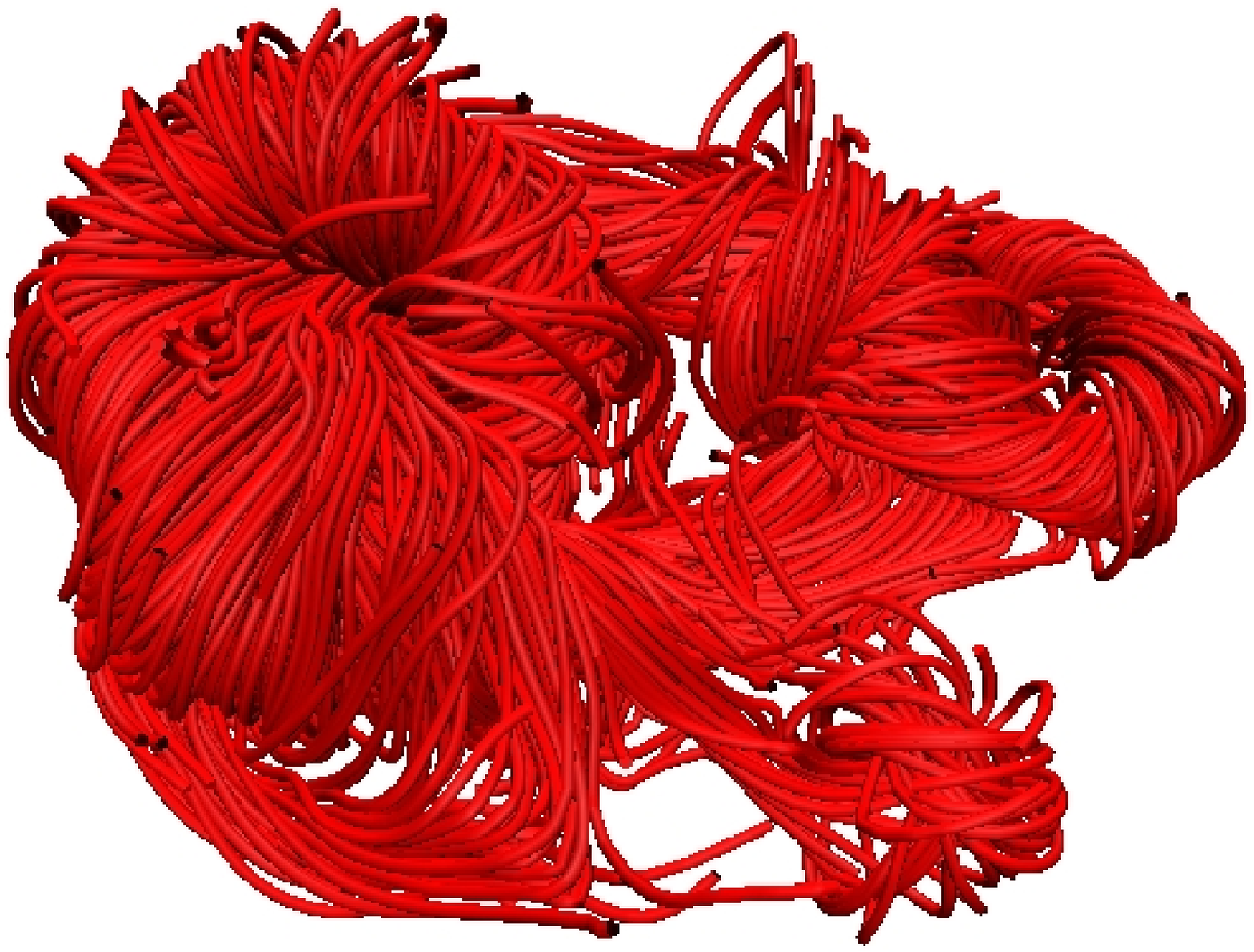}
\includegraphics[width=0.33 \hsize,angle=0]{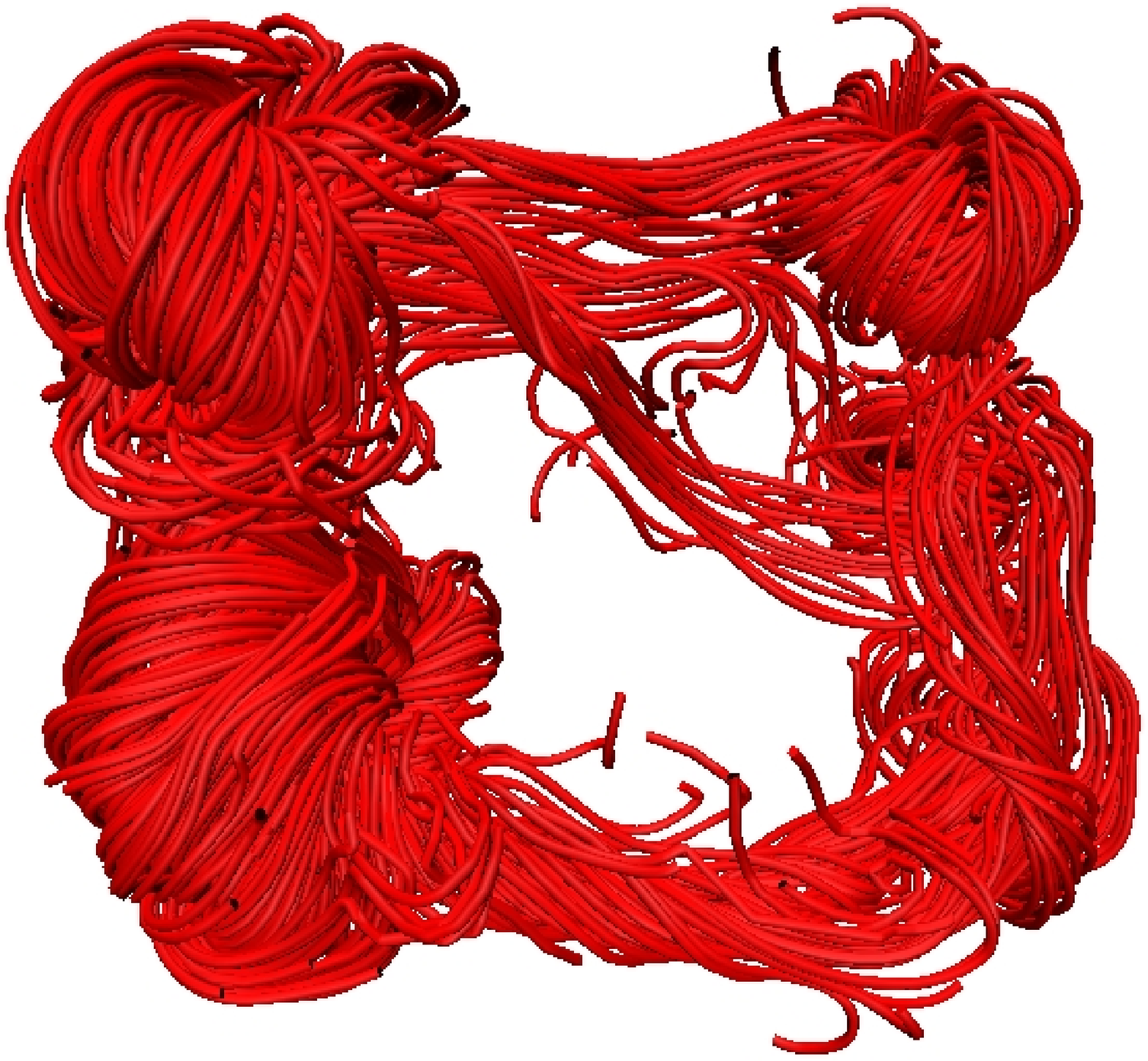}
\includegraphics[width=0.33 \hsize,angle=0]{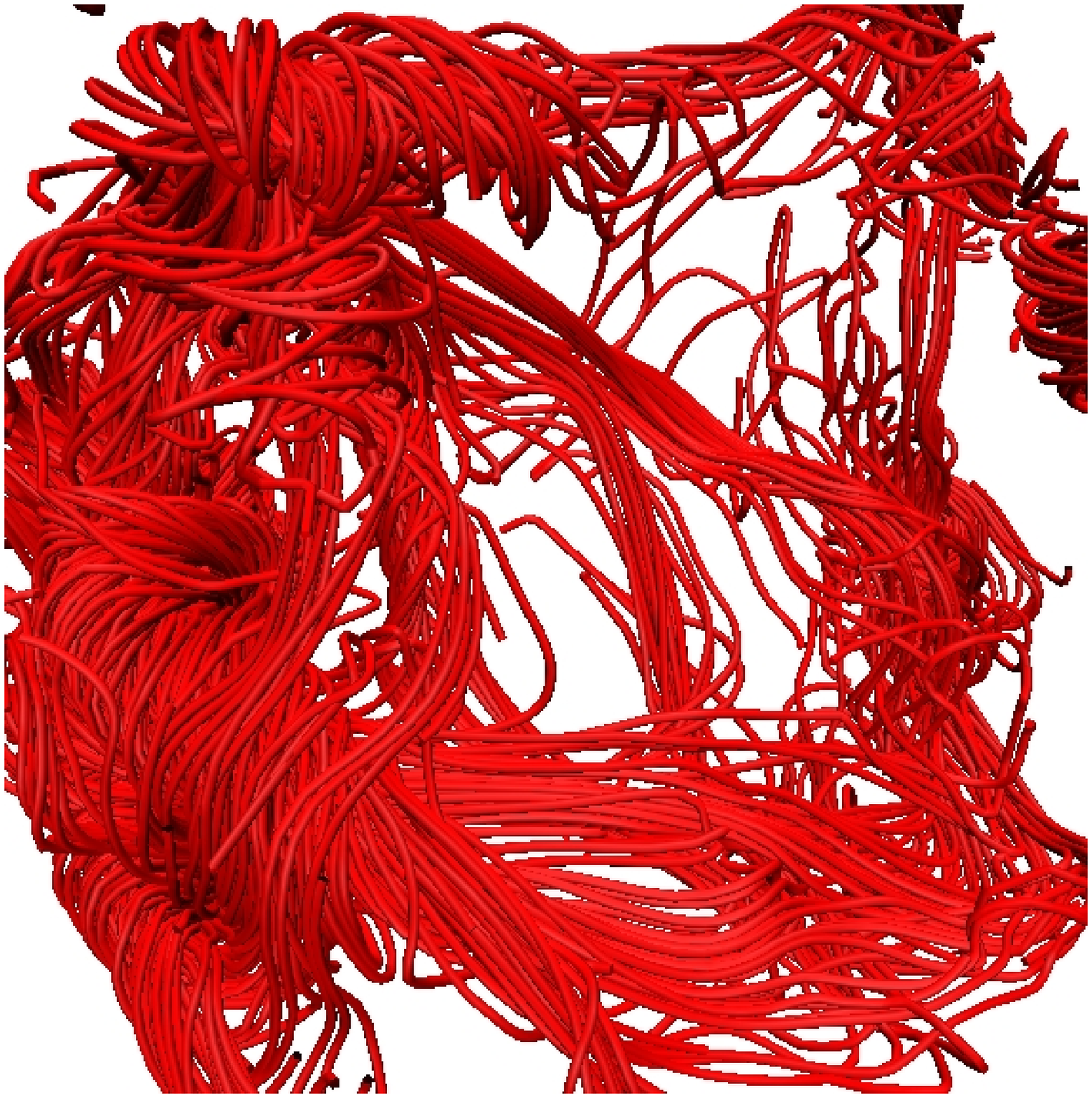}\\
\parbox[t]{0.73\hsize}{\mbox{}\\
\includegraphics[width=0.43\hsize,angle=0]{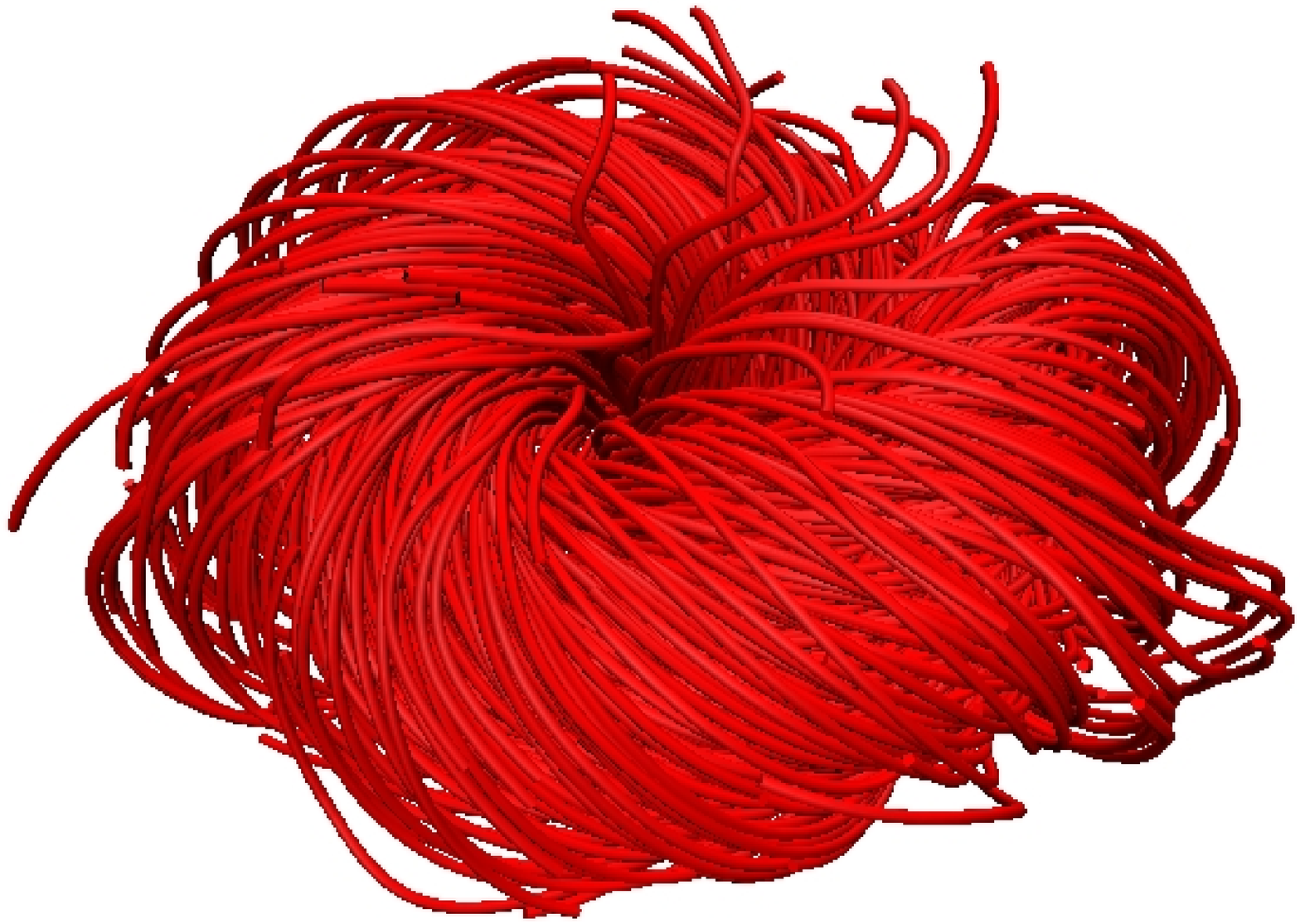}
\includegraphics[width=0.56\hsize,angle=0]{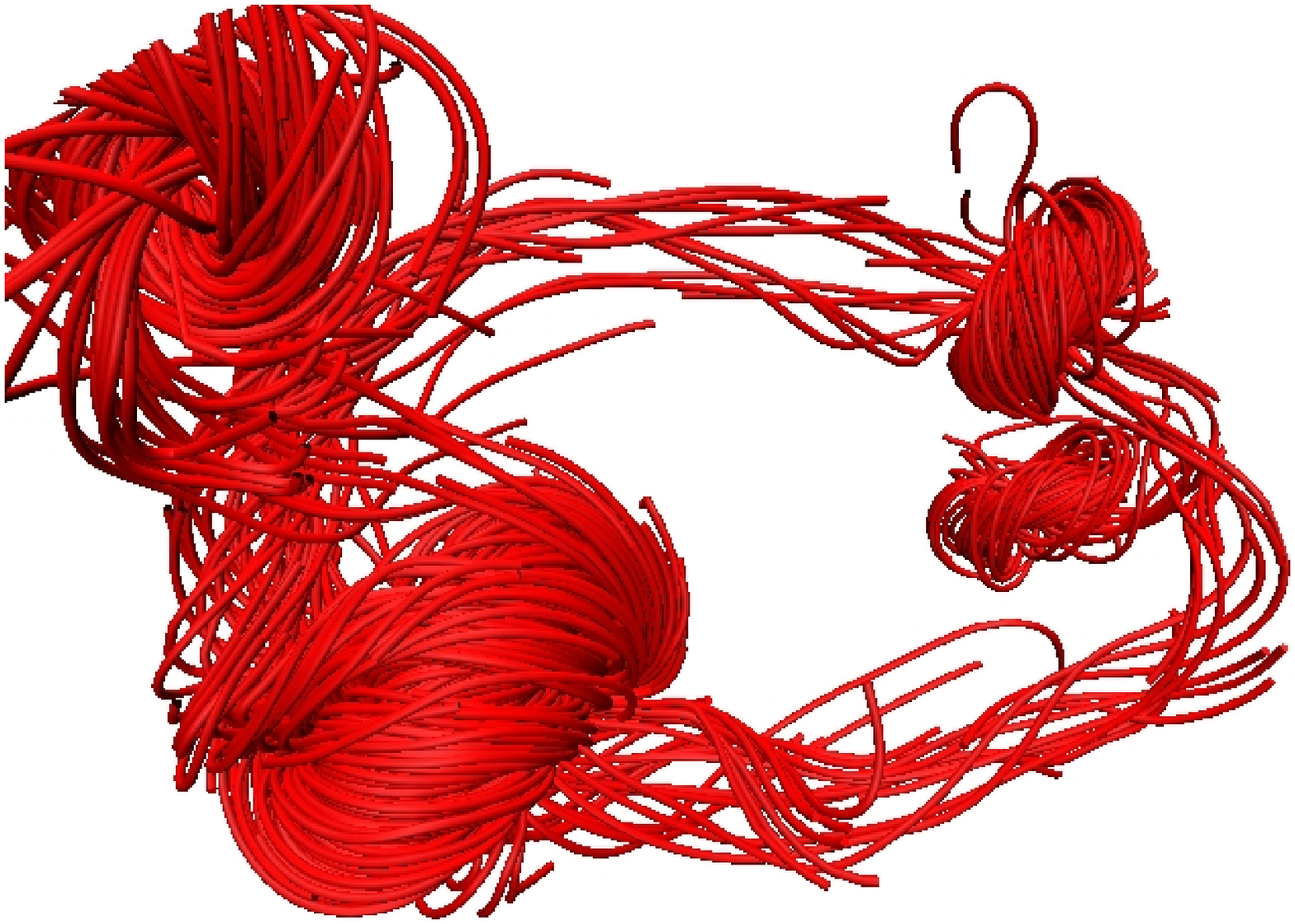}}
\parbox[t]{0.24\hsize}{
\caption{{\it Top row}: \label{fig:vapor2} contemporaneous snapshots from three otherwise identical simulations with $\rho_{\rm i}/\rho_{\rm o}=0.1$, $1$ and $10$ (left to right respectively). {\mk The $0.1$ case is run `e' in the previous section.} {\it Bottom row}: the two bubbles with $\rho_{\rm i}/\rho_{\rm o}=0.1$ and $1$ at a later time; in the highest density case the bubble spills over the boundaries of the computational box and no second snapshot is plotted.}}
\end{figure*}

We now explore the effect of the bubble density. To do this, simulations were run with the following values of the density ratio $\rho_{\rm i}/\rho_{\rm o}$ parameter: $0.1$ (as above), $1$ and $10$. Of course, the former ($0.1$) is the only ratio consistent with the observations but it is informative to look at other values. In fig.~\ref{fig:vapor2}
 three simulations are compared which have identical initial conditions except for $\rho_{\rm i}/\rho_{\rm o}$; their behaviour is quite different.

The main difference between the simulations is that the denser bubbles move into the surroundings more easily, become more non-spherical and are more prone to breaking up, even though the field strength is the same and the Alfv\'en speed is lower. In the figure we see that in the low-density case (left column) the bubble becomes somewhat distorted but then returns to a more spherical shape, forming a simple torus. In the run with $\rho_{\rm i}/\rho_{\rm o}=1$ (middle column) different parts of the bubble move away from each other and several small torus shapes are formed, connected by weak flux tubes.

\begin{figure*}
\includegraphics[width=0.28\hsize,angle=0]{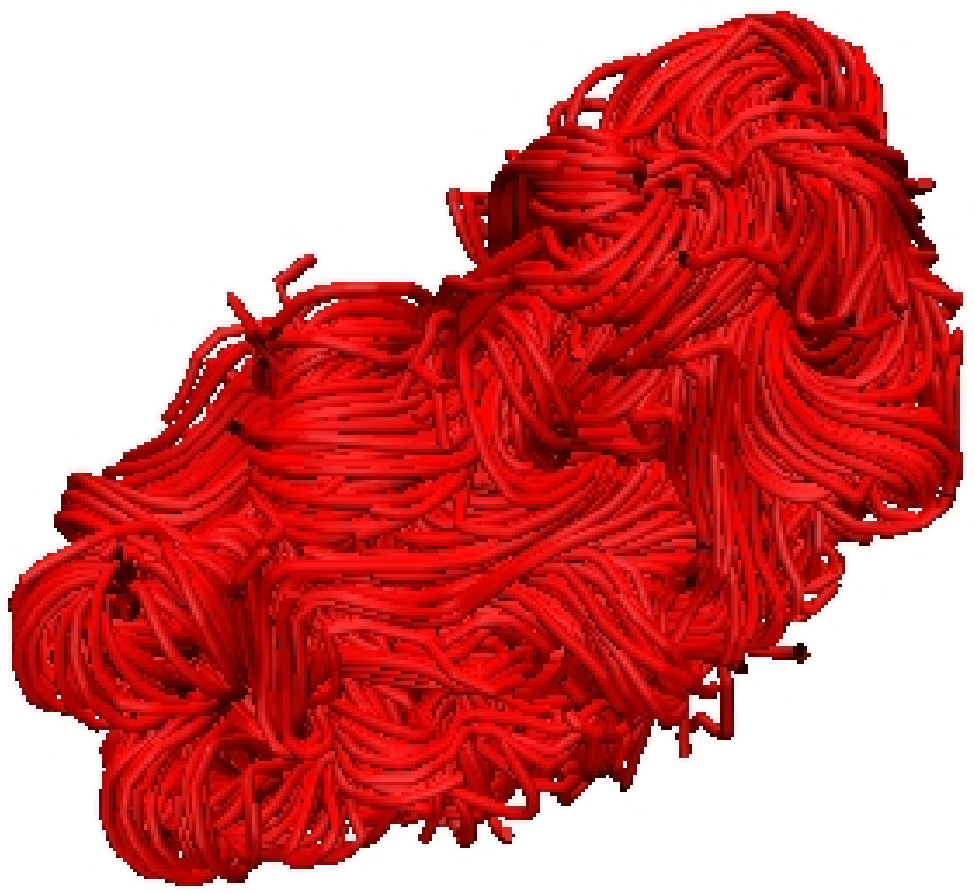}
\includegraphics[width=0.33\hsize,angle=0]{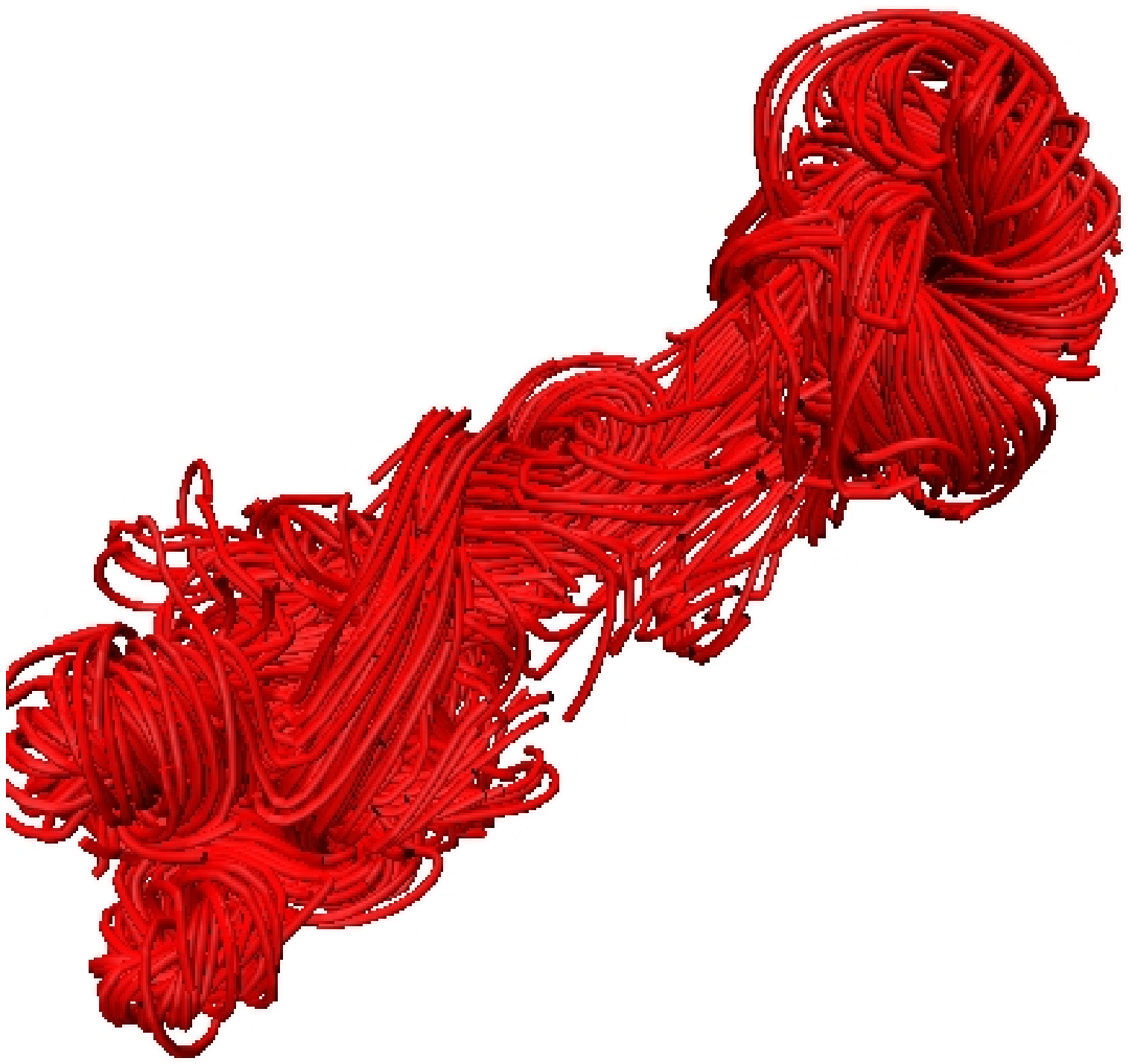}
\includegraphics[width=0.37\hsize,angle=0]{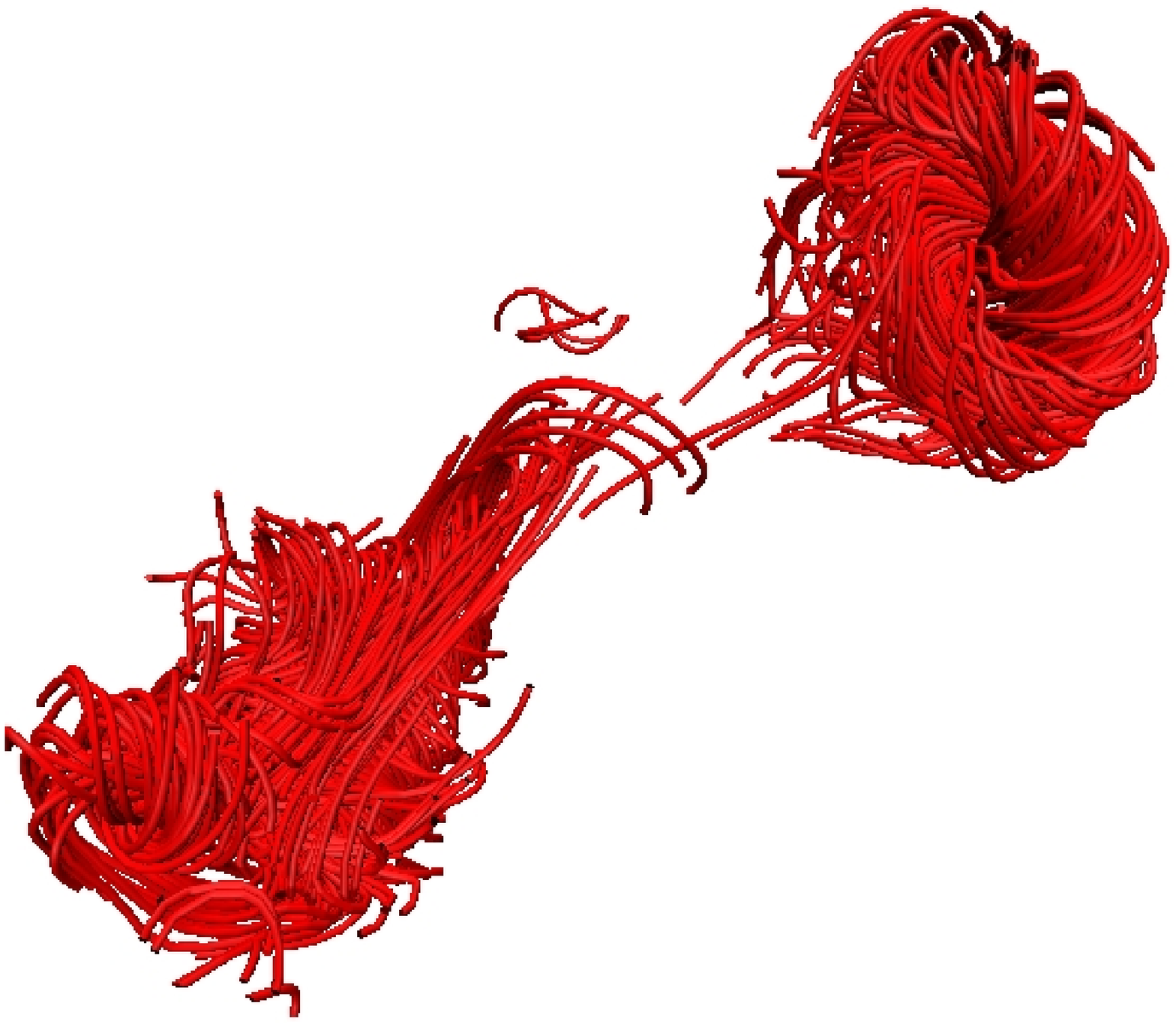}\\
\caption{Time sequence from a simulation with $\rho_{\rm i}=\rho_{\rm o}$ and a rather low helicity {\mk (the initial magnetic field is identical to run `g' in figs.~\ref{fig:fid-en-t} to \ref{fig:fid-en-hel}).} The bubble first becomes elongated and then breaks into two parts, connected only by very weak flux tubes.}
\label{fig:break_up}\end{figure*}

To illustrate how a bubble can break up, field lines of a run with $\rho_{\rm i}/\rho_{\rm o}=1$ are plotted in fig.~\ref{fig:break_up}. This run has very low helicity $\lambda_{\rm i}=0.00025$, {\mk and the magnetic field is identical to run `g' described in the previous section}. The bubbles with higher helicity always form a simple torus-shaped equilibrium regardless of the density.

This dependence on the bubble density can be understood in the following way. During reconnection, material inside the bubble is moving around with velocity comparable to the Alfv\'en speed, so that the bubble will not stay spherical for very long. The kinetic energy density of the plasma is comparable to the magnetic energy density $B^2/8\pi$ and its momentum per unit volume is $B\sqrt{ \rho/ 4\pi }$. A denser (and colder) bubble has more momentum and can penetrate the surrounding medium more easily. Alternatively, one can think of the distance over which a projectile slows down via aerodynamic drag: it is comparable to the distance over which it has to push its own mass out of the way, which is obviously further if it is more dense. Since in reality we know that bubbles have a low density, i.e. $\rho_{\rm i}/\rho_{\rm o}<1/3$ and probably even lower, we should expect only a modest deformation of the bubble from the Alfv\'enic motions inside them.

\subsection{Effect of resolution}
{\mk 
It is often useful when employing numerical methods to examine the effect of resolution. To this end, simulations were run at double the spatial resolution (i.e. half the grid-spacing $\Delta x$) used in the simulations described in the previous sections. To avoid excessive computational cost, the computational box was made smaller: the domain had sides $L=4.5r_{\rm i}$, as opposed to $6r_{\rm i}$ as used previously; in the particular simulations run, it was checked that the boundaries did not cause significant problems. To be sure of separating the effects, some low resolution simulations were run with $L=4.5r_{\rm i}$, so that grid spacings of $\Delta x=r_{\rm i}/24$ (as used previously) and $r_{\rm i}/48$ could be compared. In fig.~\ref{fig:high_res} the evolution of the energy and helicity of the magnetic field in two otherwise identical sets of simulations are presented. 

\begin{figure}
\includegraphics[width=1.0\hsize,angle=0]{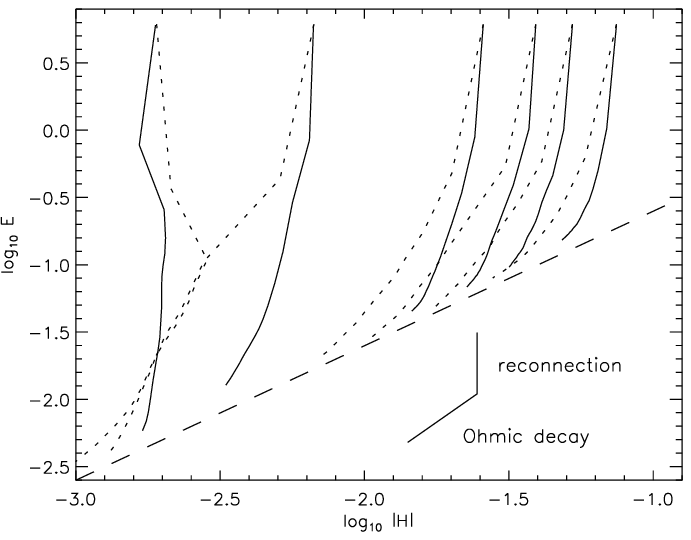}
\caption{Log energy against log helicity for two sets of simulations performed at standard resolution $\Delta x=r_{\rm i}/24$ (dotted lines) and double that resolution (solid lines). It is very clear that with improving resolution, helicity is better conserved during relaxation to equilibrium.}
\label{fig:high_res}
\end{figure}

It can be seen in the figure that helicity conservation improves at higher resolution. Physically, higher resolution is equivalent to lower diffusivity (kinetic, thermal and magnetic), and we would expect that this causes the thickness of reconnection sheets to fall. This is because the reconnection speed is set by the Alfv\'en speed and whatever microscopic (diffusive) processes are involved therefore have to work at a given rate; lower diffusivity means that the process has to take place on a smaller length scale\footnote{That the relaxation of a magnetic field into equilibrium involves current sheets at all can be seen from the same argument. A rigorous proof of the formation of current sheets in three-dimensional magnetic relaxation does not exist, however; \citet{Gruzinov:2009} discusses a proof of current sheet formation in two dimensions.}. This in turn means that less helicity is destroyed in the reconnection zones, since helicity has units of energy $\times$ length. This validates the assumption of approximate helicity conservation (\ref{eq:hel_cons}). Finally, note that increased resolution causes the equilibrium, once formed, to evolve diffusively (downwards and to the left in the fig.\ \ref{fig:high_res}) more slowly.
}

\section{Structure of the equilibria}\label{sec:structure}

In an MHD equilibrium in a medium without gravitational force, the Lorentz force $(1/4\pi)(\nabla\times{\bf B})\times{\bf B}$, which is perpendicular to the magnetic field, is balanced only by the pressure gradient force $-\nabla P$, so we can see that the magnetic field lines lie on surfaces of constant pressure. Since pressure is a scalar field, the equilibrium must consist of nested magnetic surfaces. This is clearly inaccessible from the turbulent initial conditions considered here without some topological reorganisation, i.e. reconnection, hence some loss of magnetic energy is inevitable.

We see from the simulations that the basic building block from which the various equilibria are made is the twisted flux tube; in the simplest case which I describe in more detail below, there is a single circular-looped twisted flux tube, and in other cases the twisted flux tubes are arranged in more complex patterns. I shall now explore the properties of these tubes.

\subsection{Twisted flux tubes}\label{sec:tubes}

{\mk The magnetic configurations reached from lower-helicity initial conditions in general consist of intertwined, branched, twisted flux tubes. The tension in a flux tube is proportional to $2B^2_{\rm ax}-B^2_{\rm h}$, where $B_{\rm ax}$ and $B_{\rm h}$ are the axial (parallel to tube axis) and hoop (perpendicular to tube axis) field components; see the Appendix for a proof. This means that a flux tube with only a modest twist, i.e. with $B_{\rm h}\ll B_{\rm ax}$, will have a tension, tending to shorten the tube until $2B^2_{\rm ax}=B^2_{\rm h}$. The tubes connecting the two tori in fig.~\ref{fig:merger} are connected by such untwisted flux tubes, with the result that the two tori are pulled towards each other and merge.} In contrast, the flux tubes visible in the top-centre frame of fig.~\ref{fig:vapor2} are twisted, with the result that the structures are not pulled towards one another and at a later time (lower-right frame) are still at similar distances.

A twisted flux tube will in general be subject to an interchange instability of low azimuthal wavenumber \citep{Shafranov:1956,Tayler:1957,Kruskaletal:1958}.
It is found that in the limit of low diffusivities, the tube becomes unstable to a $m=1$ `kink' mode when the number of field-line windings around the tube exceeds unity, i.e. when
\begin{equation}\label{eq:kink}
\frac{l}{a} > 2\pi\frac{B_{\rm ax}}{B_{\rm h}}.
\end{equation}
In light of the zero-tension condition $2B^2_{\rm ax}=B^2_{\rm h}$, this in effect sets an upper limit on the length of a flux tube. The instability will result in reconnection, loss of magnetic energy and hoop flux $\Phi_{\rm h}$. As $\Phi_{\rm h}$ falls, the tube will tend to contract and become wider, thus restoring equilibrium. The instability will be quenched as the left-hand-side of (\ref{eq:kink}) falls and becomes equal to the right-hand-side.

The tendency of a long flux tube to become shorter in this way is in some sense a manifestation of the tendency of a flux tube of a given helicity to reduce its energy (although it is not clear whether the instability just described does in fact conserve helicity). Now, if we allow $\Phi_{\rm ax}$ and $\Phi_{\rm h}$ to change whilst holding their product $H$ constant (as might be the situation during initial reconnection) then we have (ignoring some factors of order unity)
\begin{equation}
H\sim\Phi_{\rm ax}\Phi_{\rm h}\approx\pi a^3 l B_{\rm ax}B_{\rm h}\approx aVB^2,
\end{equation}
the last step following from the equilibrium condition (\ref{eq:rel_ax_h}). The energy of the field $E\sim VB^2$, so that
\begin{equation}
\frac{E}{H} \sim \frac{1}{a}.
\end{equation}
This means that the lowest energy state of a flux tube with a given volume and helicity is $l\rightarrow0$, $a\rightarrow\infty$. The reason that short, fat flux tubes do not actually appear in the simulations presumably has something to do with the boundaries at the ends of the tubes and with limits on the flux $\Phi_{\rm ax}$ through these boundaries. One can imagine various kinds of end-boundaries for a flux tube, the simplest being periodic, or in other words, the tube is connected to itself in a loop. In this case, it is clear that the length $l\ge2\pi a$, and although of course the assumption of a straight tube with circular cross-section made above will no longer be accurate, one would certainly expect to find a tendency for tube loops to contract as much as possible and reach $l\approx2\pi a$ (via kink instability once a tube has formed, or otherwise); this case where a flux tube loop is a circle is discussed in the next section. Also possible is that the tube merges with other tubes (some of which might resemble simple circular loops), perhaps in such a way that the ends of the tube may be considered anchored and that $\Phi_{\rm ax}$ is fixed. This kind of tube can be seen in the top-centre and top-right frames of fig.~\ref{fig:vapor2} where various flux tubes are attached to each other in various ways.

One flux-tube arrangement which often makes an appearance in the simulations is the figure-of-eight and variations thereof. In its simplest form it consists of a single flux tube wrapped around itself in a double-helix fashion. Its total length is apparently greater than the kink-instability limit given by (\ref{eq:kink}) and (\ref{eq:rel_ax_h}), which could be allowed because this configuration is in fact the end-state of the instability or because there is some degree of branching, with field lines passing between neighbouring sections of the tube. Whilst this question and the detailed structure of this configuration is left for future investigation, it is possible to understand intuitively why the tube should twist around itself in this manner. Imagine starting with a straight flux tube with given axial flux  but with zero twist, i.e. $B_{\rm h}=0$. Now rotate one end of the tube by an angle $\alpha$, and note that $B_{\rm h}\propto\alpha$ and that the hoop-component energy $E_{\rm h}\propto\alpha^2$ so that the torque required to produce an additional rotation $\delta\alpha$ is proportional to $\alpha$. Now imagine two parallel tubes -- both with $B_{\rm h}=0$ -- touching along their length and joined somehow at the ends (so strictly speaking, one flux tube). Now rotating one end of the configuration by an angle $\beta$ about the axis created by the line of contact of the two tubes will produce a new field component perpendicular to that axis and it is easily verified that the torque required to produce an additional rotation is again proportional to the angle $\beta$. Finally, imagine a configuration where the two tubes have an initial $B_{\rm h}$ as if they had been twisted by an angle $\alpha$; the tubes will exert a torque on the ends in such a way that a helical rotation will be produced in the opposite direction from the internal twist of the field lines inside the tubes, until the two torques are in balance. Since the tubes are touching and therefore the radii of rotation are comparable, in the end state $\alpha\sim\beta$. This approximate relation is consistent with the simulations.

It is not clear exactly how the transition is made from the double-helix configuration to the simple torus, although it is common in the simulations (see fig.~\ref{fig:vapor1}). However, there should be some drop in hoop flux in relation to axial flux. A deeper investigation of this point is left to the future.

\subsection{Simple axisymmetric equilibrium}\label{sec:axisym}

The most basic self-contained equilibrium is that shown in the right-hand frame of fig.~\ref{fig:vapor1}. This configuration is approximately symmetrical about some axis.\footnote{A way of predicting the direction of this axis from the initial conditions has so far escaped investigation.} In an axisymmetric equilibrium the azimuthal component of the Lorentz force must vanish, leading to the condition \citep{Mestel:1961}:
\begin{equation}\label{eq:mestel61}
{\bf B}_{\rm p}\cdot{\bm\nabla}(\varpi B_\phi)=0,
\end{equation}
where ${\bf B}_{\rm p}$ and $B_\phi$ are the poloidal (meridional) and toroidal (azimuthal) components of the magnetic field and $\varpi$ is the cylindrical radius. Everything can be described in terms of a flux function $\psi(\varpi,z)$ which is the product of the cylindrical radius and the azimuthal component of the vector potential ${\bf A}$, i.e. $\psi=\varpi A_\phi$ where ${\bf B}_{\rm p}={\bm\nabla}\times(A_\phi {\bf e}_\phi)$, so that
\begin{eqnarray}
\varpi B_z = \frac{\partial \psi}{\partial \varpi} \;\;\;\;\;\;{\rm and}\;\;\;\;\;\; \varpi B_\varpi = -\frac{\partial \psi}{\partial z};\\
P=P(\psi) \;\;\;\;\;\;\;\;{\rm and}\;\;\;\;\;\;\;\; \varpi B_\phi=F(\psi).\label{eq:PandF}
\end{eqnarray}
The condition that the Lorentz and pressure gradient forces balance can now be expressed
\citep{GraandRub:1958,Shafranov:1966} as:
\begin{eqnarray}\label{eq:grad-shafranov}
& &\!\!\!\!\!\!\!\!\!\!\!\! \Delta^* \psi + \frac{1}{2}\frac{{\rm d}(F^2)}{{\rm d}\psi} + 4\pi\varpi^2\frac{{\rm d}P}{{\rm d}\psi}=0,\\
& & \;\;\;\;\;{\rm where}\;\;\;\;\;\;\;\;\;\;\; \Delta^*\psi\equiv \frac{\partial^2\psi}{\partial z^2} + \varpi \frac{\partial}{\partial w} \left(\frac{1}{\varpi}\frac{\partial\psi}{\partial\varpi}\right)\nonumber.
\end{eqnarray}
Note that (\ref{eq:mestel61}) and (\ref{eq:grad-shafranov}) are the equivalents of (\ref{eq:no_ax_force}) and (\ref{eq:tube-gs-eqn}) in this geometry; see the Appendix for details.

\begin{figure}
\includegraphics[width=1.0\hsize,angle=0]{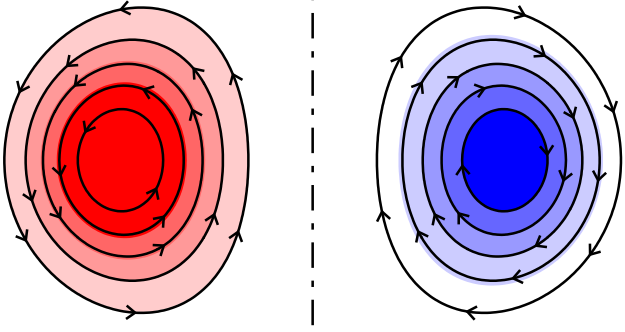}
\caption{Contour plot of $\psi$ in the meridional $(\varpi,z)$ plane in the equilibrium illustrated on the right-hand-side of fig.~\ref{fig:vapor1}. From the definition of $\psi$, the contours are also poloidal field lines, whose direction is indicated by the arrows. The contours are equally spaced, meaning that there is equal poloidal flux between each of the plotted field lines. [Note that one consequence of this is that the density of lines plotted vanishes towards the axis, although the poloidal field is strong in this central region.] In addition, the left and right hand sides contain contour plots of pressure difference $P-P_{\rm o}$ and of $F=\varpi B_\phi$ respectively, represented by red and blue shading. It is clear that $P=P(\psi)$ and $F=F(\psi)$ as predicted above.} 
\label{fig:psi}
\end{figure}

It is informative now to look in more detail at the torus equilibria found in the simulations. Once a conversion to a suitable cylindrical coordinate system using the axis of symmetry of the equilibrium, quantities can be plotted in the $(\varpi,z)$ plane. In fig.~\ref{fig:psi} we see contours of the flux function $\psi$ as well as of $F$ and $P$ in the highest-helicity simulation of the set described in section \ref{sec:results_hel}, with $\lambda_{\rm i}=0.032$; clearly the contours coincide, as in (\ref{eq:PandF}). We can also see that the cross-section of the flux tube is slightly elliptical close to its axis (the `neutral line' where the poloidal field, or ${\bf B}_{\rm h}$, vanishes) and a more complex shape further away; this is simply the tube's response to the torus geometry.

\begin{figure}
\includegraphics[width=1.0\hsize,angle=0]{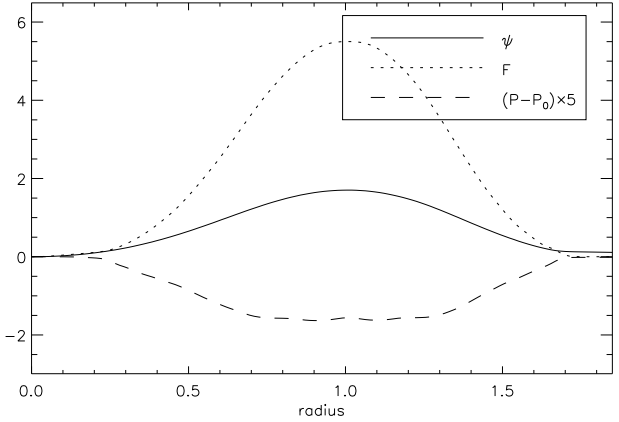}
\caption{Plots of $\psi$, $F$ and $P-P_{\rm o}$ (multiplied for clarity by $5$)
 against cylindrical radius $\varpi$ along the line $z=0$ in the meridional plane, in the equilibrium illustrated in fig.~\ref{fig:psi}.
}\label{fig:z0}
\end{figure}

In fig.~\ref{fig:z0} various quantities are plotted in the equatorial plane $z=0$. Clearly, $F(\psi)$ is a smoother function than $P(\psi)$; it is possible that fluctuations present during the formation of the torus have become `frozen' into the equilibrium. Despite these fluctuations, the field is confirmed to satisfy (\ref{eq:grad-shafranov}) to better than one part in $100$ throughout the volume.

If the energies in the three components of the magnetic field ($B_\varpi,B_\phi,B_z$) are calculated as fractions of the total magnetic energy, the values ($0.172,0.497,0.331$) are found. Whether the proximity of these fractions to $1/6,1/2,1/3$ is coincidence is a matter of speculation at this stage. In the case of a {\it straight} flux tube, the fractional energies are $1/3,1/3,1/3$ (see the Appendix). In addition, it is found that if the quantity $P_{\rm mag}\equiv B^2/24\pi$ is averaged over the entire volume, it is found to be within 2\% of the average of the mean gas pressure difference $P_{\rm o}-P$. That the average internal gas pressure is lower than the external pressure is a general result; the bubble cannot have a gas pressure excess confined by magnetic tension, an idea which is often found in the literature. It is however conceivable that a configuration exists where the gas pressure is higher than the external pressure at some region within the bubble, although its mean must still be lower than the external pressure.

\section{Discussion}
\label{sec:disc}

The equilibria found in the simulations had low field strength, i.e. high plasma-$\beta$. A low-$\beta$ plasma would drive shocks into the external medium and could be numerically rather inconvenient. While there should be no fundamental difference between the structure of equilibria with $\beta\gg1$ and those with $\beta\approx1$, since the pressure gradient force in both cases is able to balance the Lorentz force, an equilibrium with $\beta\ll1$ should look rather different -- it must be approximately force-free inside the bulk of the bubble; the pressure term drops out of (\ref{eq:grad-shafranov}). In addition there must be a Lorentz force singularity on the boundary balanced by a discontinuity\footnote{In practice the `discontinuity' should have some finite width, perhaps even non-negligible in comparison to the size of the bubble.} in gas pressure between the bubble interior and the external medium. Since the structure of such an equilibrium is constrained to a greater extent than that of a non-force-free equilibrium, it should be easier to construct such a field using analytic methods. Numerical investigation of this kind of equilibrium is left for the future. {\mk Note that the inhibition of the Rayleigh-Tayler instability at the bubble's leading edge might require a magnetic pressure comparable to the pressure of the ICM, meaning that we should expect disruption of a high-beta bubble. \citet{DeYoung:2003} discusses growth timescales and stabilising field strengths.}

Also left for the future is a more thorough study of the helicity and energy input from the AGN outflow. In section \ref{sec:jet} we produced the greatest possible helicity by having the accretion disc and resultant jet threaded by a net flux, but it is not at all certain whether this should be the case in reality. The accretion disc can carry flux inwards from the source cloud to the outflow launching region \citep{SprandUzd:2005} but it is unclear what magnetic field geometry the source is likely to have. It is plausible that the net flux through the launching region fluctuates in time faster than a bubble inflates, so that the bubble contains regions with magnetic helicity of opposite signs. This will presumably mean that the first step once reconnection begins will be formation of localised equilibria (as in the high-helicity case), followed by mutual annihilation and finally the formation of an extremely low-energy global equilibrium, if there is sufficient time. Another possibility is that the accretion does not have a significant net flux and the launching region has mixed magnetic polarity. In this case, the helicity of the bubble is extremely low, even in small regions; it is not understood how the field should evolve in this case but it seems likely that reconnection should dissipate almost all of the magnetic energy on a rather short timescale.

We have assumed here that there is no mixing of bubble material with the external medium. While the possibly of external material entering the bubble by {\it diffusive} processes is ruled out by \citet{Pope:2010}, it is possible that some external material becomes entrained in the bubble. First, there is the possibility of entrainment while the bubble is being inflated; the jet transfers momentum to the rest of the bubble in a potentially irregular manner and external material could easily become advected inwards. \citet{Croston_etal:2008} for instance find evidence that in low-power radio sources (FR-I objects, \citealt{FanandRil:1974}), the pressure difference between the external medium and the radio-emitting component of the cavities is much more likely to be {\mk entrained} external material than relativistic protons; the possibility of a strong magnetic field however remains. In addition, during the reconnection process there are blobs of material moving around at the Alfv\'en speed (which may exceed the sound speed in the external medium) inside the bubble; it is plausible that some external matter becomes advected into the bubble. This material could become a non-magnetised `island' inside the bubble or pass through a reconnection region and become magnetised. Conversely, bubble material could leave the bubble as magnetised islands, but would have to pass through a reconnection region in order to become magnetically disconnected from the rest of the bubble.

It has also been assumed here that the gas in the bubble is at rest before reconnection begins. In reality, the gas is fired into the bubble at high speed and loses most of its kinetic energy in a shock but some will be left over, resulting in flows inside the bubble. Neglecting this motion simplified the analysis somewhat, but one can speculate on its possible effect. The fluid viscosity in such a diffuse medium is rather high so that such motions will be damped on a short timescale; to be more precise, $\tau_{\rm visc}=l^2/\nu\sim l^2/(\lambda c_{\rm s})$ where $\nu$, $\lambda$ and $c_{\rm s}$ are the viscosity, mean free path and sound speed, respectively. With realistic parameters we find that the viscous timescale is comparable to the sound-crossing time. This is almost certainly less than the buoyant rise timescale calculated in (\ref{eq:risetime}). However while these motions persist, they may prevent relaxation to equilibrium if the kinetic energy is greater than the magnetic energy. This is unlikely in the case of a magnetically-accelerated outflow where only some fraction of the magnetic energy is converted to kinetic.

{\mk As pointed out by \citet{Pfrommeretal:2005}, in} the future it should be possible to use the Sunyaev-Zel'dovich (S-Z) effect to measure some of the unknown parameters (\citealt{SunandZel:1972}; for a review see \citealt{Birkinshaw:1999}; {\mk see \citealt{Basuetal:2010} for the most recent results).}  Whereas the X-ray brightness is a line-of-sight integral, roughly speaking, of the square of the gas density, the S-Z intensity is a line-of-sight integral of the gas pressure. It may or may not be that we shall see dark cavities in the S-Z effect -- it will be possible therefore to distinguish between a magnetically-dominated bubble ($\beta\ll1$) and a gas-pressure-dominated bubble ($\beta\gg1$). In addition, by measuring the spectrum of the effect it is possible to distinguish between a non-relativistic and a relativistic plasma, which will resolve the issue of thermal gas vs.\ cosmic ray pressure.

\subsection{Other astrophysical contexts}

The equilibria found here show some similarity to those thought to reside in various kinds of non-convective star: upper-main-sequence, white dwarfs and neutron stars. A star which starts its life with some chaotic accretion/convection phase and a turbulent magnetic field undergoes the same kind of relaxation to equilibrium once the convection dies away. The main difference between a star and a bubble is that in the former, gravity restricts motion in the radial direction. This provides extra stability, so that a greater range of stable equilibria is possible. For instance, it is possible in a star to have a much larger toroidal component than poloidal component \citep{Braithwaite:2009} whereas the two must be roughly equal in a bubble. Another way of thinking about this is that there are two forces -- the pressure gradient $-{\bm\nabla}P$ and gravity $\rho{\bf g}$ -- which can be adjusted independently of each other to balance the Lorentz force, which also has two degrees of freedom (remember that ${\bm\nabla}\cdot{\bf B}=0$ removes one degree).

There is also a striking similiarity between the twisted flux tubes found in the simulations and structures found in the ionosphere of Venus. The Pioneer spacecraft measured the magnetic field vector as it orbited the planet, frequently encountering regions of strong magnetic field. \citet{RusandElp:1979} found that the spacecraft was flying through twisted flux tubes. The tubes, about $10$km wide and with field strengths around $400\,\mu$G, have a magnetic pressure comparable to the external gas pressure, although it is not clear whether the gas pressure in the tubes is lower than or comparable to the magnetic pressure, i.e. whether the tubes have $\beta\ll 1$ or $\beta\sim 1$. The surrounding ionosphere is much more weakly magnetised, with $B\approx 20\,\mu$G. Similar tubes have since been found in the ionospheres of Mars \citep{Cloutier_etal:1999} and Titan. From the considerations in section \ref{sec:tubes}, it should not be surprising that the tubes are twisted, as that state is in some sense the `natural' condition of a flux tube. However, any flux tube needs to be held at the ends, either by a torque (in the case of a twisted tube with $2B_{\rm ax}^2=B_{\rm h}^2$) or by a tension (in the case of an untwisted tube).

\section{Conclusions}
\label{sec:conc}

I have considered the evolution of the magnetic field inside AGN-inflated bubbles which are observed as dark cavities in X-ray images of galaxy clusters. It is found that the magnetic field undergoes relaxation to a global-scale equilibrium filling the entire bubble, consisting of twisted flux tube(s) arranged in some pattern. The relaxation process inevitably involves magnetic reconnection -- the reconnection regions could provide energetic synchrotron-emitting particles via X-point and Fermi acceleration {\mk(see e.g. \citealt{Parker:1957,Miller_etal:1997}).}

The timescale on which this relaxation takes place, or in other words the stage during this relaxation we are likely to observe, depends crucially on various parameters: the magnetic field strength, mass density, Lorentz factor and size of the outflow as well as the properties of the ambient intra-cluster medium into which the bubble expands. Given the uncertainly in these parameters, it is impossible at this stage to distinguish between the following eventualities (see section \ref{sec:timescales}). In the following, the radius of the bubble and the dominant length scale of its magnetic field structure are $r$ and $l$ respectively; the Alfv\'en speed is $v_{\rm A}=B/\sqrt{4\pi\rho}$ and there is a reconnection timescale $\tau_{\rm rec}=l/(\alpha v_{\rm A})$ where $\alpha\approx0.1$ is the reconnection speed parameter. 
\begin{enumerate}
\item The AGN outflow is weakly magnetised and little reconnection occurs; the observed field is small-scale and evolves passively in response to the bubble's interaction with the ICM. Measurement of the relevant parameters would show that $l\ll r$ and $\tau_{\rm rec}>\tau_{\rm age}$ where $\tau_{\rm age}$ is the age of the bubble.
\item The AGN outflow is strongly magnetised and the magnetic field relaxes towards a global equilibrium. However, because the helicity of the field is low, the bulk of the magnetic energy is dissipated and no global equilibrium is reached. At the time of observation the reconnection is still ongoing and $\tau_{\rm rec} \approx \tau_{\rm age}$. The field may consist of local-equilibrium flux tubes of size $l<r$.
\item The AGN outflow is strongly magnetised and has high helicity so that reconnection proceeds quickly and a global equilibrium is reached: $l\approx r$ and $\tau_{\rm rec} > \tau_{\rm age}$. The field may consist of large-scale twisted flux tube(s) arranged in figure-of-eight patterns or as a single torus configuration, similar to the spheromak shape found in laboratory experiments. In this case, the magnetic field will give the bubble some rigidity, helping keep it intact as it moves through the ICM.
\end{enumerate}
To illustrate this with plausible parameters, if we measure a density $10^{-5}m_p$g cm$^{-3}$ and field strength $20\mu$G in a bubble of radius $10$kpc then $\tau_{\rm rec}\approx (l/r)\,7$Myr; if the bubble is older than $7$Myr then we have the global-equilibrium case (iii). To reach this situation the AGN outflow must have had high helicity;
this is likely if the accretion disc is fed material with a consistent net flux. Fluctuating or vanishing net flux through the accretion disc will result in case (ii) even if the outflow is strongly magnetised.

During reconnection to equilibrium, the shape of the bubble may change in response to plasma flow inside the bubble on the order of the Alfv\'en speed. However, if the density of the bubble is much less than the density of the surrounding ICM, the effect on the shape of the bubble will be rather modest.

Finally, it is shown that the difference in gas pressure between a bubble and its surroundings is equal to one third of the magnetic energy density, i.e. the magnetic field produces an `isotropic magnetic pressure' $P_{\rm mag}=B^2/24\pi=P_{\rm o}-P_{\rm i}$ where the subscripts o and i denote pressure outside and inside the magnetised volume. In this and other contexts this is more useful than the $B^2/8\pi$ which is more common in the literature; this is a general feature of three-dimensional problems.

{\it Acknowledgements.} The author would like to thank Marcus Br\"uggen, Eugene Churazov, Peter Goldreich, \AA ke Nordlund, Christoph Pfrommer and Henk Spruit for assistance and useful discussions.

\begin{appendix}
\section{Structure and tension of a flux tube}

Consider a straight flux tube, uniform along its length, with some arbitrary cross-section. The field component parallel to the direction of the tube is $B_{\rm ax}$, whilst ${\bf B}_{\rm h}$ represents the other two components which can be thought of as field lines in the plane of the cross-section.\footnote{The suffices ax and h stand for `axial' and `hoop'. The reader will see that finding a consistent terminology for both this case and the case of the axisymmetric equilibria described in section \ref{sec:axisym} is less than straightforward. In the literature, normally $z$ and $\phi$ are used for the axial and hoop directions respectively. However, when a tube is connected into a circular loop the axial direction becomes `azimuthal' and is denoted by $\phi$, and $z$ is now the axis of the loop.} We can see from the argument in section \ref{sec:structure} that in equilibrium the pressure $P$ is constant on lines of ${\bf B}_{\rm h}$. Furthermore, the axial component of the Lorentz force must vanish in a tube which is uniform along its length, since there is no axial pressure gradient to balance it. From this it can be shown that:
\begin{equation}
{\bf B}_{\rm h}\!\cdot\!\nabla B_{\rm ax}=0,
\label{eq:no_ax_force}\end{equation}
which in other words means that $B_{\rm ax}$ must be constant along lines of ${\bf B}_{\rm h}$. We can now describe the magnetic and pressure fields in terms of the axial component $A_{\rm ax}$ of the vector potential:
\begin{equation}
{\bf B}_{\rm h}={\bm\nabla}\times(A_{\rm ax}{\bf e}_{\rm ax}); \;\;\;P=P(A_{\rm ax});
\;\;\; B_{\rm ax}=B_{\rm ax}(A_{\rm ax});
\end{equation}
where ${\bf e}_{\rm ax}$ is the axial unit vector. Equating the Lorentz force to the pressure gradient gives
\begin{equation}\label{eq:tube-gs-eqn}
\nabla^2A_{\rm ax} + \frac{1}{2}\frac{{\rm d}(B_{\rm ax}^2)}{{\rm d}A_{\rm ax}} + 4\pi\frac{{\rm d}P}{{\rm d}A_{\rm ax}}=0.
\end{equation}
The three equations (\ref{eq:no_ax_force}) to (\ref{eq:tube-gs-eqn}) are the equivalent of equations (\ref{eq:mestel61}) to (\ref{eq:grad-shafranov}) in this geometry.

In addition to this, there is a good reason to believe that tubes will have a circular cross-section. Using cylindrical coordinates ($r,\phi,z$), imagine perturbing a tube with circular cross section, radius $r_0$ and field ${\bf B}_{\rm h}\!=\!{\bf e}_\phi B_{\phi 0}(r)$ with a displacement field ${\bm \xi} = {\bf e}_r r x(\phi)$ where  $r$, ${\bf e}_r$, $\phi$ and ${\bf e}_\phi$ are the radial and azimuthal coordinates and unit vectors and $x$ is an arbitrary function. If $r'=r+\xi=(1+x)r$ then ${\rm d}r'=(1+x){\rm d}r$ and the magnetic energy per unit length of the $B_{\rm h}$ component is
\begin{eqnarray}
E_{\rm h} \!\!\!&=&\!\!\! \frac{1}{8\pi}\! \int^{\phi=2\pi}_{\phi=0} \!\!\int^{r=r_0}_{r=0} r' {\rm d}r' {\rm d}\phi\, [B_\phi^2+B_r^2]\\
\!\!\!&=&\!\!\! \frac{1}{8\pi}\! \int^{\phi=2\pi}_{\phi=0} \!\!\int^{r=r_0}_{r=0}(1+x)^2 r\,{\rm d}r\,{\rm d}\phi\, \left[ \left(\frac{B_{\phi 0}}{1+x}\right)^2 + B_r^2 \right],\nonumber
\end{eqnarray}
where $B_\phi=B_{\phi 0}/(1+x)$ follows from flux freezing. Clearly the energy in the $\phi$ component of the field is unchanged by the perturbation but for any function $x$ which is not independent of $\phi$, i.e. that gives the tube a non-circular cross-section, there is new energy in the $B_r$ component which was absent before, meaning the total energy has increased. Circular tubes therefore represent an energy minimum. The first term in the equilibrium condition (\ref{eq:tube-gs-eqn}) simplifies to $(1/r)({\rm d}/{\rm d}r)(r\,{\rm d}A_{\rm ax}/{\rm d}r)$.

Now consider such a tube with circular cross-section of radius $a$, length $l$ and volume $V=\pi a^2 l$ containing r.m.s. axial and azimuthal field components $B_{\rm ax}$ and $B_{\rm h}$. First I examine the energetics and stability of the tube to various pertubations; later I look at the boundaries at either end of the tube.

The axial and hoop fluxes $\Phi_{\rm ax}$ and $\Phi_{\rm h}$ of the magnetic field are given by
\begin{equation}
k_{\rm ax}\Phi_{\rm ax}=\pi a^2 B_{\rm ax} \;\;\;\;\;\;\;\;{\rm and}\;\;\;\;\;\;\;\; k_{\rm h}\Phi_{\rm h}=a l B_{\rm h},
\label{eq:tube_fluxes}\end{equation}
where $k_{\rm ax}$ and $k_{\rm h}$ are dimensionless factors of order unity which correct for the different averaging required when calculating energy and flux.\footnote{Instead of using averages one could consider a particular flux tube, for instance with functions $B_{\rm ax}(r)={\rm const}$ and $B_{\rm h}(r)=B_0r/a$, but this provides no extra insight.} The fluxes are conserved on a dynamical timescale, as is the magnetic helicity\footnote{Helicity is gauge independent if the domain is bounded by magnetic surfaces and/or periodic conditions. Here, we have the former on the sides of the tube and can assume the latter for the ends.} of the tube which can be expressed as $H\sim \Phi_{\rm ax}\Phi_{\rm h}$. The magnetic energy of the tube is given by
\begin{eqnarray}\label{eq:mag_en}
E\!\!\!&=&\!\!\!\frac{V}{8\pi}(B_{\rm ax}^2 + B_{\rm h}^2)
\\\!\!\!&=&\!\!\!\frac{1}{8}\left(\frac{l^2}{\pi V} k_{\rm ax}^2\Phi_{\rm ax}^2+\frac{1}{l}k_{\rm h}^2\Phi_{\rm h}^2\right)
\\\!\!\!&=&\!\!\!\frac{1}{8}\left(\frac{V}{\pi^3 a^4} k_{\rm ax}^2\Phi_{\rm ax}^2+\frac{\pi a^2}{V}k_{\rm h}^2\Phi_{\rm h}^2\right).
\end{eqnarray}
Furthermore, as the two fluxes are constant during dynamic adjustments the derivatives w.r.t. $V$ at constant $l$ and $a$ are
\begin{eqnarray}\label{eq:ender1}
\left(\frac{\partial E}{\partial V}\right)_l &=& -\frac{l^2k_{\rm ax}^2\Phi_{\rm ax}^2}{8\pi V^2}=-\frac{B_{\rm ax}^2}{8\pi},\\
\left(\frac{\partial E}{\partial V}\right)_a &=& \frac{k_{\rm ax}^2\Phi_{\rm ax}^2}{8\pi^3a^4} - \frac{\pi a^2k_{\rm h}^2\Phi_{\rm h}^2}{8V^2}=\frac{B_{\rm ax}^2}{8\pi}-\frac{B_{\rm h}^2}{8\pi}\label{eq:ender2}\end{eqnarray}
respectively, using (\ref{eq:tube_fluxes}) and specifying that the derivatives represent homogeneous expansion where $k_{\rm ax}$ and $k_{\rm h}$ are constant. Now defining a magnetic pressure given by $P_{\rm mag}=-{\rm d} E/{\rm d} V$, we see from (\ref{eq:ender1}) that a tube of fixed length provides a positive pressure in the lateral direction. Therefore once a dynamical equilibrium has been reached the average thermal pressure in the tube $P_{\rm i}$ should be lower than that in the surroundings $P_{\rm o}$ by a quantity
\begin{equation}\label{eq:lat_eq}
P_{\rm mag}=P_{\rm o}-P_{\rm i}=\frac{B_{\rm ax}^2}{8\pi}.
\end{equation}
The same must be true in the axial direction, as the thermal pressure acts the same in both directions and so therefore must the Lorentz force. This means that the derivatives (\ref{eq:ender1}) and (\ref{eq:ender2}) are equal, giving
\begin{equation}
B_{\rm h}^2 = 2B_{\rm ax}^2.
\label{eq:rel_ax_h}\end{equation}
It is also possible to imagine a tube which has reached equilibrium in the lateral direction so that relation (\ref{eq:lat_eq}) is satisfied, but which for reasons to do with whatever it is anchored to at the ends, is {\it not} in equilibrium in the axial direction. In other words, the tube has a net tension or pressure along its length. This tension is calculated thus:
\begin{equation}
T=\left[\left(\frac{\partial E}{\partial V}\right)_a + P_{\rm o}-P_{\rm i}\right] \pi a^2 = (2B_{\rm ax}^2-B_{\rm h}^2)a^2/8,
\end{equation}
using (\ref{eq:ender2}) and (\ref{eq:lat_eq}). The term $P_{\rm o}-P_{\rm i}$ comes from the fact that in stretching the tube, work $P_{\rm o}\,{\rm d}V$ must be done against the external medium while the internal gas does work $P_{\rm i}\,{\rm d}V$. Alternatively, to avoid doing $P\,{\rm d}V$ work the tube may be stretched at constant volume, in which case the tension $T=(\frac{\partial E}{\partial l})_V$, which gives the same result. Of course, the result cannot depend on the change in $a$ during the stretching, because lateral force balance has already been assumed and changes in $a$ are energetically neutral. Also note that the equilibrium condition (\ref{eq:rel_ax_h}) corresponds to vanishing tension $T$.\footnote{\citet{Weiss:1964} and authors of various later works erroneously state the tension of a flux tube as $(B_{\rm ax}^2-B_{\rm h}^2)a^2/8$, neglecting the $P\,{\rm d}V$ work.} Finally, note that as assumed in (\ref{eq:pressure}) we have
\begin{equation}\label{eq:pmag}
P_{\rm mag}=\frac{B^2}{24\pi},
\end{equation}
where $B$ is the total magnetic field given by $B^2=B_{\rm ax}^2+B_{\rm h}^2$.

\label{lastpage}
\end{appendix}

\bsp

\end{document}